\documentclass[twocolumn,showpacs,aps,prd,floatfix]{revtex4}
\usepackage{graphicx}
\usepackage{dcolumn}
\usepackage{amsmath}
\usepackage{epsfig}
\usepackage{multirow}

\newcommand{\BABARPubYear}{12}
\newcommand{\BABARPubNumber}{002}
\newcommand{\SLACPubNumber}{14957}

\input babarsym

\newcommand{\gevcccc}{\ensuremath{{\mathrm{\,Ge\kern -0.1em V^2\!/}c^4}}\xspace}

\def\Kmaybestar {\ensuremath{K^{(*)}\xspace}}

\def\kll {\B\to\Kmaybestar\ellell\xspace}
\def\kmaybeee {\B\to\Kmaybestar\epem\xspace}

\def\kmaybemm {\B\to\Kmaybestar\mumu\xspace}

\def\mkpi {\ensuremath{m_{\kaon\pi}}\xspace}

\def\modeeight {\ensuremath{B^0\rightarrow \Kp \pim \mumu}\xspace}
\def\modeeightshort {\ensuremath{\Kp \pim \mumu}\xspace}
\def\modeeleven {\ensuremath{B^+\rightarrow \KS \pip \epem}\xspace}
\def\modeelevenshort {\ensuremath{\KS \pip \epem}\xspace}

\def\modefour {\ensuremath{B^+\rightarrow K^+\epem}\xspace}
\def\modefourshort {\ensuremath{K^+\epem}\xspace}

\def\modekavgll {\ensuremath{B\to K\ellell}\xspace}

\def\modekmaybell {\B\to\Kmaybestar\ellell\xspace}

\def\modekstll {\ensuremath{B\rightarrow K^{*}\ellell}\xspace}

\def\modeone {\ensuremath{B^0\rightarrow \KS \mumu}\xspace}
\def\modeoneshort {\ensuremath{\KS \mumu}\xspace}
\def\modeseven {\ensuremath{B^+\rightarrow \KS \pip \mumu}\xspace}
\def\modesevenshort {\ensuremath{\KS \pip \mumu}\xspace}

\def\modethree {\ensuremath{B^0\rightarrow \KS \epem}\xspace}
\def\modethreeshort {\ensuremath{\KS \epem}\xspace}
\def\modetwelve {\ensuremath{B^0\rightarrow \Kp \pim \epem}\xspace}
\def\modetwelveshort {\ensuremath{\Kp \pim \epem}\xspace}
\def\modetwo {\ensuremath{B^+\rightarrow K^+\mumu}\xspace}
\def\modetwoshort {\ensuremath{K^+\mumu}\xspace}

\long\def\inst#1{\par\nobreak\kern 4pt\nobreak
    {\it #1}\par\vskip 10pt plus 3pt minus 3pt}

\def\etal{{\it et al.}}

\begin{document}

\begin{flushleft}
SLAC-PUB-\SLACPubNumber \\
\babar-PUB-\BABARPubYear/\BABARPubNumber
\end{flushleft}

\title{
\large \bf
\boldmath
         Measurement of Branching Fractions and Rate Asymmetries in the Rare Decays $\kll$
} 

%
\author{J.~P.~Lees}
\author{V.~Poireau}
\author{V.~Tisserand}
\affiliation{Laboratoire d'Annecy-le-Vieux de Physique des Particules (LAPP), Universit\'e de Savoie, CNRS/IN2P3,  F-74941 Annecy-Le-Vieux, France}
\author{J.~Garra~Tico}
\author{E.~Grauges}
\affiliation{Universitat de Barcelona, Facultat de Fisica, Departament ECM, E-08028 Barcelona, Spain }
\author{A.~Palano$^{ab}$ }
\affiliation{INFN Sezione di Bari$^{a}$; Dipartimento di Fisica, Universit\`a di Bari$^{b}$, I-70126 Bari, Italy }
\author{G.~Eigen}
\author{B.~Stugu}
\affiliation{University of Bergen, Institute of Physics, N-5007 Bergen, Norway }
\author{D.~N.~Brown}
\author{L.~T.~Kerth}
\author{Yu.~G.~Kolomensky}
\author{G.~Lynch}
\affiliation{Lawrence Berkeley National Laboratory and University of California, Berkeley, California 94720, USA }
% ...
\collaboration{The \babar\ Collaboration}
\noaffiliation


\begin{abstract}
   \noindent
In a sample of 471 million $\BB$ events collected with the \babar\, detector at the \pep2\ $\epem$ collider we study the rare decays $\kll$, where $\ell^+\ell^-$ is either $e^+e^-$ or $\mu^+\mu^-$.
We report results on partial branching fractions and isospin asymmetries in seven bins of di-lepton mass-squared.
We further present $\CP$ and lepton-flavor asymmetries for di-lepton masses below and above
the $\jpsi$ resonance. We find no evidence for $\CP$ or lepton-flavor violation. The partial branching fractions 
and isospin asymmetries are consistent with the Standard Model predictions
and with results from other experiments.
\end{abstract}


\maketitle


\section{Introduction}
\label{sec:introduction}

\begin{figure}[b!]
\begin{center}
\includegraphics[height=3cm]{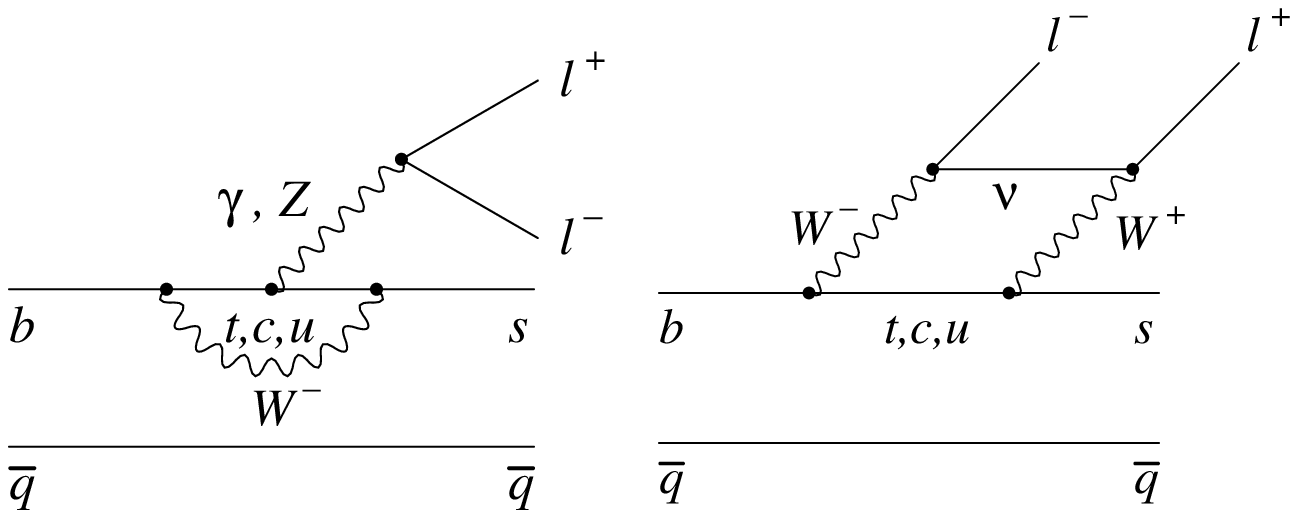}
\caption{Lowest-order Feynman diagrams for $\btosll$.}
\label{fig:sll_diagrams}
\end{center}
\end{figure}

The decays $\kll$ arise from flavor-changing neutral-current processes
that are forbidden at tree level in the Standard Model (SM).
The lowest-order SM processes contributing to these decays
are the photon penguin, the $Z$ penguin and the $W^+W^-$ box diagrams
shown in Fig.~\ref{fig:sll_diagrams}. Their amplitudes are expressed
in terms of hadronic form factors and perturbatively-calculable
effective Wilson coefficients, $C^{\rm eff}_{7}$,  $C^{\rm eff}_{9}$ and $C^{\rm eff}_{10}$,
which represent the electromagnetic penguin diagram,
      and the vector part and the axial-vector part of the linear combination of the $Z$ penguin and $W^+W^-$ box diagrams, respectively~\cite{Buchalla}. In next-to-next-to-leading order (NNLO) at a renormalization scale
$\mu=4.8$~\gev,
    the effective Wilson coefficients are $C^{\rm eff}_7 =-0.304$,  $C^{\rm eff}_9 =4.211$, and
 $C^{\rm eff}_{10} =-4.103$~\cite{Altmannshofer:2008dz}.

\begin{figure}[t!]
\begin{center}
\includegraphics[width=0.32\linewidth]{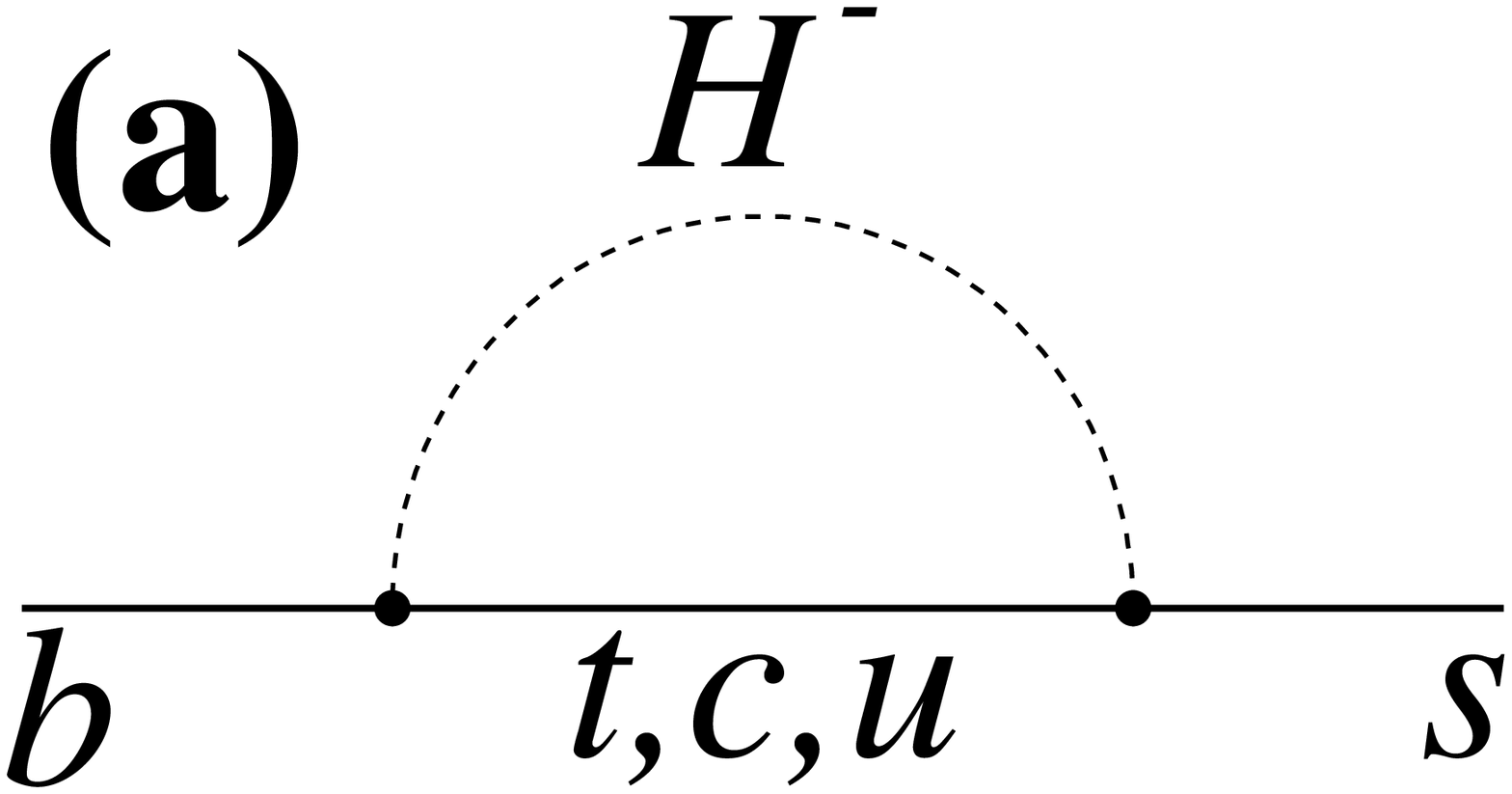}
\includegraphics[width=0.32\linewidth]{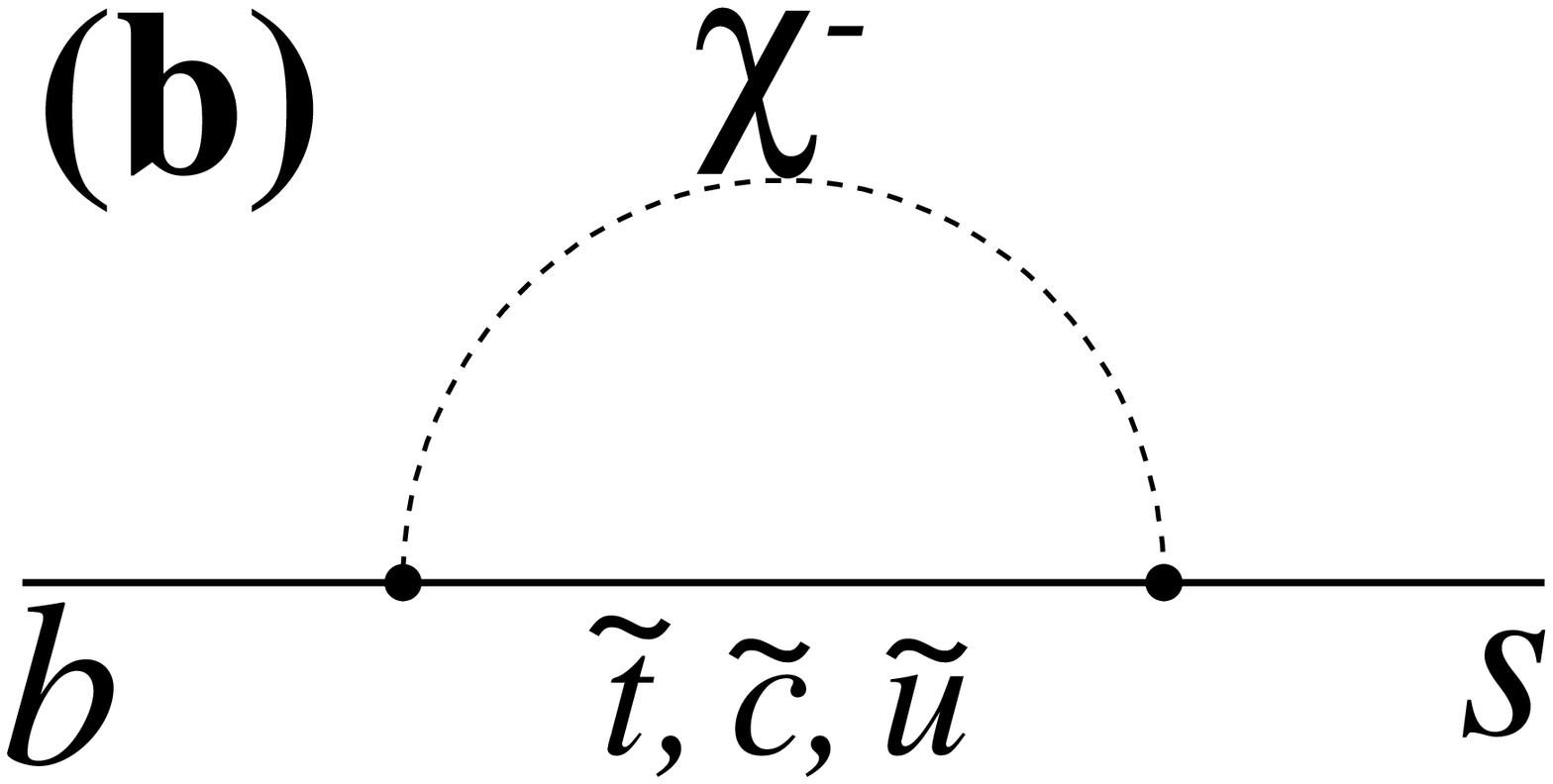}
\includegraphics[width=0.32\linewidth]{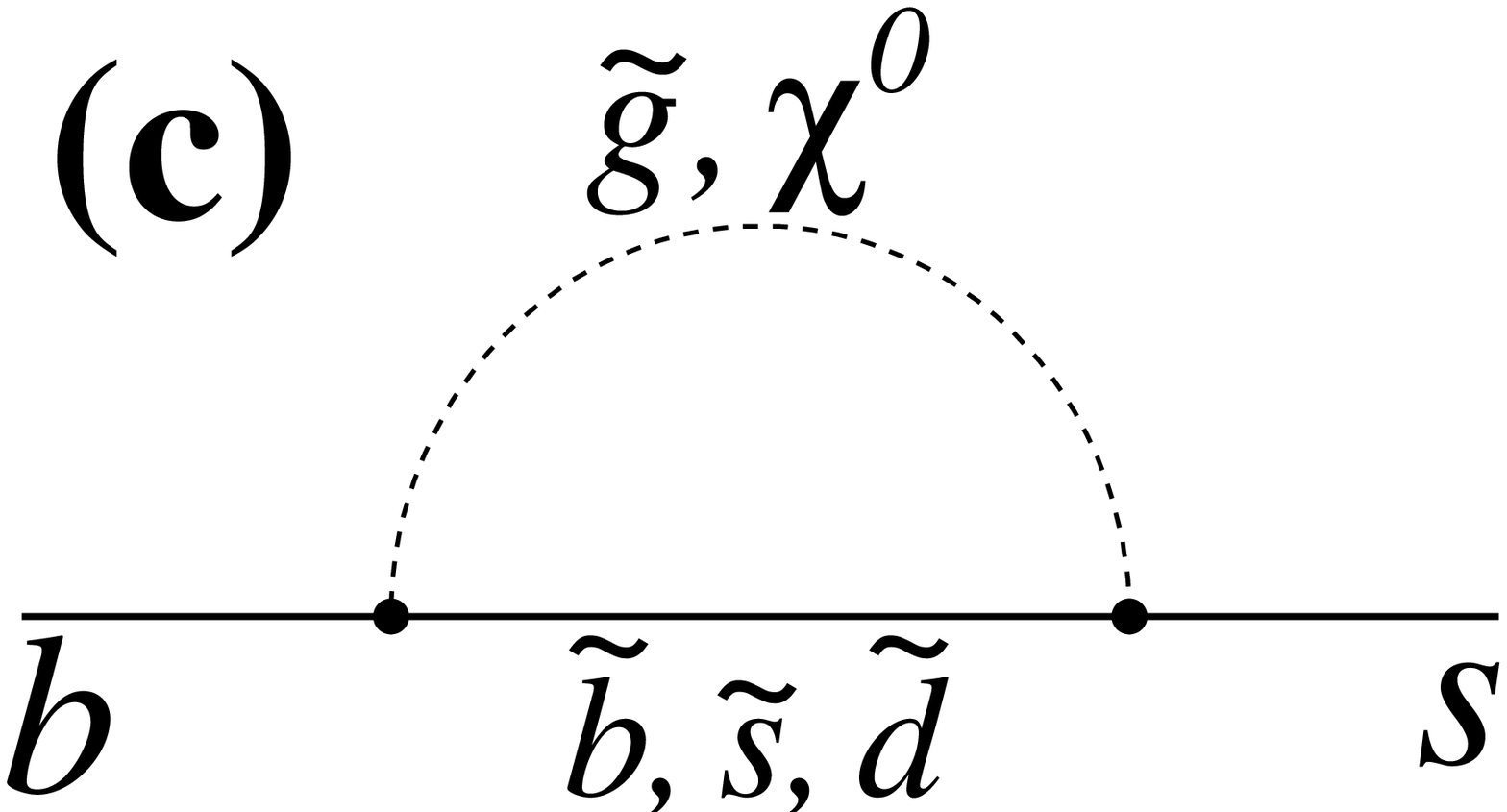}
\caption{Examples of new physics loop contributions to $\btosll$: (a) charged Higgs ($H^-$); (b) squark ($\tilde{t}, \tilde{c}, \tilde{u}$) and chargino ($\chi^-$); (c) squark ($\tilde{b}, \tilde{s}, \tilde{d}$) and gluino ($\tilde{g}$)/neutralino ($\chi^0$).}
\label{fig:sll_np}
\end{center}
\end{figure}

Non-SM physics may add new penguin and box diagrams, which can 
contribute at the same order as the SM diagrams~\cite{NewPhysics, isospin, Ali:2002jg}. Examples of new physics
loop processes are depicted in Fig.~\ref{fig:sll_np}. These contributions might modify the Wilson coefficients from
their SM expectations~\cite{Ali:2002jg, Zhong:2002nu, Ali01, Khodjamirian:2010vf}. In addition, new contributions from scalar,
      pseudoscalar, and tensor currents may arise that can modify, in particular, the lepton-flavor ratios~\cite{Alok:2010zd,Yan:2000dc}.

\section{Observables}
\label{sec:observables}

\begin{table}[b!]
\centering
  \caption[$s$ bins]{
    \label{tab:sbins}
    The definition of seven $s$ bins used in the analysis.
    Here $m_B$ and $m_{\Kmaybestar}$ are the invariant
    masses of $B$ and $\Kmaybestar$, respectively. The low $s$ region 
    is given by
  $0.10<s<8.12$~\gevcccc, while the high $s$ region is given by $s>10.11$~\gevcccc. }

  \begin{tabular}{cccc}
    \\ \hline \hline \\
   & $s$ bin & $s$ min & $s$ max \\
   & & (\gevcccc) & (\gevcccc) \\ \hline
   Low & $s_1$ & 0.10  & 2.00            \\
   & $s_2$ & 2.00  & 4.30           \\
   & $s_3$ & 4.30 & 8.12       \\\hline
  High & $s_4$ & 10.11 & 12.89    \\
   & $s_5$ & 14.21 & 16.00     \\
   & $s_6$ & 16.00  & $(m_B-m_{\Kmaybestar})^{2}$ \\ \hline
   & $s_0$ & 1.00  & 6.00            \\\hline
    \hline
  \end{tabular}

\end{table}

We report herein results on exclusive partial branching fractions and isospin asymmetries in six bins of $s\equiv m^2_{\ell \ell}$, 
defined in Table~\ref{tab:sbins}.
We further present results in the $s$ bin $s_0=1.0-6.0~{\rm GeV}^2/c^4$ 
chosen for
calculations inspired by soft-collinear effective theory (SCET)~\cite{scet}. In addition, we report on direct \CP asymmetries and  the ratio of rates to dimuon and dielectron final states
in the low $s$
    and high $s$
    regions separated by the $J/\psi$ resonance.
We remove regions of the long-distance contributions around the $J/\psi$ and $\psitwos$  resonances.
New \babar\, results on angular observables using the same dataset and similar event selection will be reported shortly.

The $B \rightarrow K \ell^+ \ell^-$ and $B \rightarrow K^* \ell^+ \ell^-$ total branching fractions are predicted to be $(0.35\pm0.12) \times 10^{-6}$ and $(1.19\pm0.39) \times 10^{-6}$ (for~$s>0.1~{\rm GeV}^2/c^4$), respectively~\cite{Ali:2002jg}. The $\sim 30\%$ uncertainties are due to a lack of knowledge about the form factors that
model the hadronic effects  in the $B\rightarrow K$ and $B\rightarrow K^*$ transitions.
 Thus, measurements of decay rates to exclusive final states are less suited to searches for new physics
than rate asymmetries, where many theory uncertainties
cancel.

For charged $\B$ decays and neutral $\B$ decays flavor-tagged
        through $K^*\to K^+\pi^-$~\cite{chargeconj},
    the direct $\CP$ asymmetry 
is defined as
\begin{eqnarray}
{\cal A}_{\CP}^{\Kmaybestar} \equiv
\frac
{{\cal B}(\overline{B} \rightarrow \overline{K}^{(*)}\ellell) - {\cal B}(B \rightarrow K^{(*)}\ellell)}
{{\cal B}(\overline{B} \rightarrow \overline{K}^{(*)}\ellell) + {\cal B}(B \rightarrow K^{(*)}\ellell)}\,,
\end{eqnarray}
\noindent and is expected to be ${\cal O}(10^{-3})$ in the SM.
However ${\cal A}_{\CP}^{\Kmaybestar}$ may receive a significant enhancement
from new physics contributions at the electro-weak scale~\cite{CPnp}.

For $s > 0.1$~\gevcccc, the ratio of rates to dimuon and dielectron final states
is defined as

\begin{eqnarray}
{\cal R}_{\Kmaybestar} \equiv
\frac
{{\cal B}(\kmaybemm)}
{{\cal B}(\kmaybeee)}\,.
\end{eqnarray}

\noindent In the SM, ${\cal R}_{\Kmaybestar}$
is expected to be unity to within a few percent~\cite{Hiller:2003js}
for dilepton invariant masses above the dimuon kinematic threshold.
In two-Higgs-doublet models, including supersymmetry, these
ratios are sensitive to the presence of a neutral Higgs
boson. 
When the ratio of  neutral Higgs field vacuum expectation values $\tan \beta$ is large,
    ${\cal R}_{K^{(*)}}$ might be increased by up to
10\%~\cite{Yan:2000dc}. 

The \CP-averaged isospin asymmetry is defined as
\begin{equation}
\resizebox{0.95\linewidth}{!}{$
{\cal A}^{K^{(*)}}_{I} \equiv
\frac
{{\cal B}(\Bz \to K^{(*)0}\ellell) - r_\tau {\cal B}(\Bp \to K^{(*)+}\ellell)}
{{\cal B}(\Bz \to K^{(*)0}\ellell) + r_\tau {\cal B}(\Bp \to K^{(*)+}\ellell)}$,
}
\label{eq:aidef}
\end{equation}
\noindent
where $r_\tau \equiv \tau_{\Bz}/\tau_{\Bp}=1/(1.071\pm 0.009)$
is the ratio of $B^0$ and $B^+$ lifetimes~\cite{PDG}. ${\cal A}^{K^{*}}_{I}$
has a SM expectation of $+6$\% to $+13$\% as $s \ra 0$~\cite{isospin}. This is consistent with the measured  asymmetry $3\pm3\%$ in $B \rightarrow K^* \gamma$~\cite{kstg}. A calculation of the predicted $K^{*+}$ and $K^{*0}$ rates integrated over the low $s$ region yields ${\calA}^{K^*}_I =-0.005\pm 0.020$~\cite{Beneke, Feldmann:ckm2008}. In the high $s$ region, we may expect
contributions from charmonium states as an additional source of isospin asymmetry. However the  measured asymmetries in
the $J/\psi K^{(*)}$ and $\psitwos\Kmaybestar$ 
modes are all below 5\%~\cite{PDG}.

\section{BABAR Experiment and Data Sample}
\label{sec:experiment}

We use a data sample of 471 million $B \bar B$ pairs collected at the $\Upsilon(4S)$ resonance with the \babar\, detector~\cite{BaBarDetector}
at the PEP-II asymmetric-energy $e^+ e^-$ collider at the SLAC National Accelerator Laboratory.
Charged particle tracking is provided by a five-layer silicon vertex tracker and a 40-layer drift chamber
in a 1.5 T solenoidal magnetic field. We identify electrons with a CsI(Tl) electromagnetic calorimeter, and muons
using an instrumented magnetic flux return. Electron and muon candidates are required to have momenta
$p > 0.3 \gevc$
 in the laboratory frame. We combine up to three photons with electrons
when they are consistent with bremsstrahlung,
and do not use electrons that are associated with photon
conversions to low-mass $\epem$ pairs.
We identify charged kaons using a detector of internally reflected
Cherenkov light, as well as $\dedx$ information from the drift chamber.
Charged tracks other than identified $e$, $\mu$ and $K$ candidates are treated as pions.
Neutral $\KS \to \pip \pim$ candidates are required to have an invariant mass
consistent with the nominal $K^0$ mass, and a flight distance
from the $\epem$ interaction point that is more than three times
its uncertainty.

\section{Event Selection}
\label{sec:selection}

We reconstruct $\modekmaybell$ signal events in the following
eight final states:
\begin{itemize}
\item $\Bz\to \KS\mumu$,
\item $\Bp\to \Kp\mumu$,
\item $\Bz\to \KS\epem$,
\item $\Bp\to \Kp\epem$,
\item $\Bp\to \Kstarp(\to \KS\pip)\mumu$,
\item $\Bz\to \Kstarz(\to \Kp\pim)\mumu$,
\item $\Bp\to \Kstarp(\to \KS\pip)\epem$,
\item $\Bz\to \Kstarz(\to \Kp\pim)\epem$.
\end{itemize}
\noindent
We reconstruct $\KS$ candidates in the $\pip\pim$ final state.
We also study the $K^{(*)} h^\pm \mu^\mp $ final states, where $h$ is a charged track with no particle identification requirement applied,
to characterize backgrounds from hadrons misidentified as muons.
We use a $K^{*} e^\pm \mu^\mp $ sample
to model 
the combinatorial background from two random leptons.
In each mode, we utilize the kinematic variables $\mes=\sqrt{E^2_{\rm CM}/4 -p^{*2}_B}$ and
$\Delta E = E_B^* - E_{\rm CM}/2$, where $p^*_B$ and $E_B^*$ are the $B$ momentum and energy in the $\Upsilon(4S)$ center-of-mass (CM) frame,
    and $E_{\rm CM}$ is the total CM energy.

For masses $\mes > 5.2$~\gevcc we perform one-dimensional fits of the $\mes$ distribution for $K\ellell$ modes. 
For $K^*\ellell$ modes, 
we include in addition the $K \pi$ mass region  $0.72 < \mkpi <1.10$~\gevcc
in the fit. 
We use the sideband $5.20 < \mes < 5.27$~\gevcc to characterize combinatorial background shapes and normalizations. For both the $e^+e^-$ and $\mu^+ \mu^-$ modes, we veto the $J/\psi (2.85 < m_{\ell \ell} < 3.18~{\rm GeV}/c^2)$ and $\psitwos (3.59 < m_{\ell \ell} < 3.77~ {\rm GeV}/c^2)$ mass regions. The vetoed events provide high-statistics control samples 
that we use 
to validate the fit methodology.

\begin{figure}
\begin{center}

\includegraphics[width=0.45\textwidth]{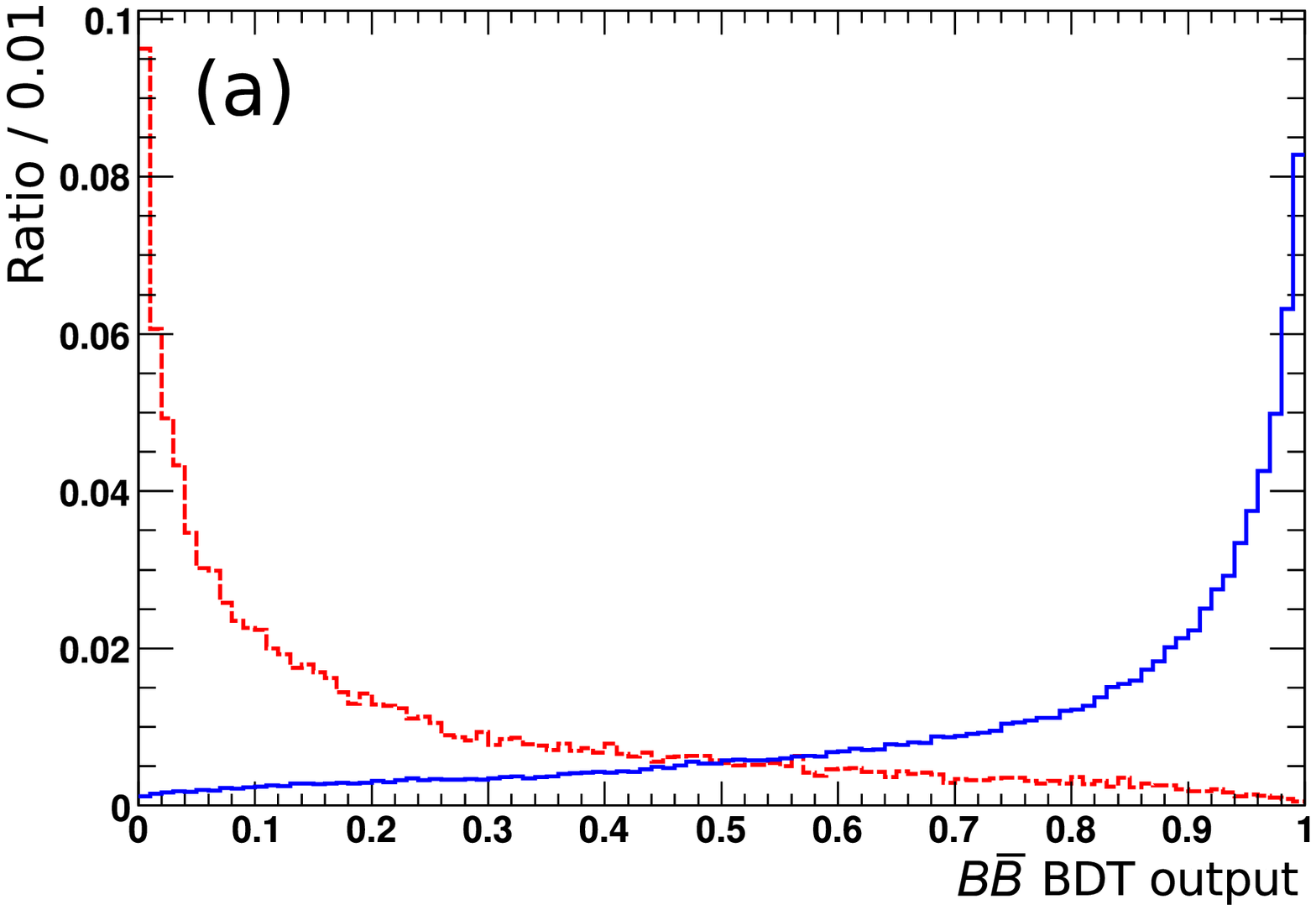}\\\vspace{5mm}
\includegraphics[width=0.45\textwidth]{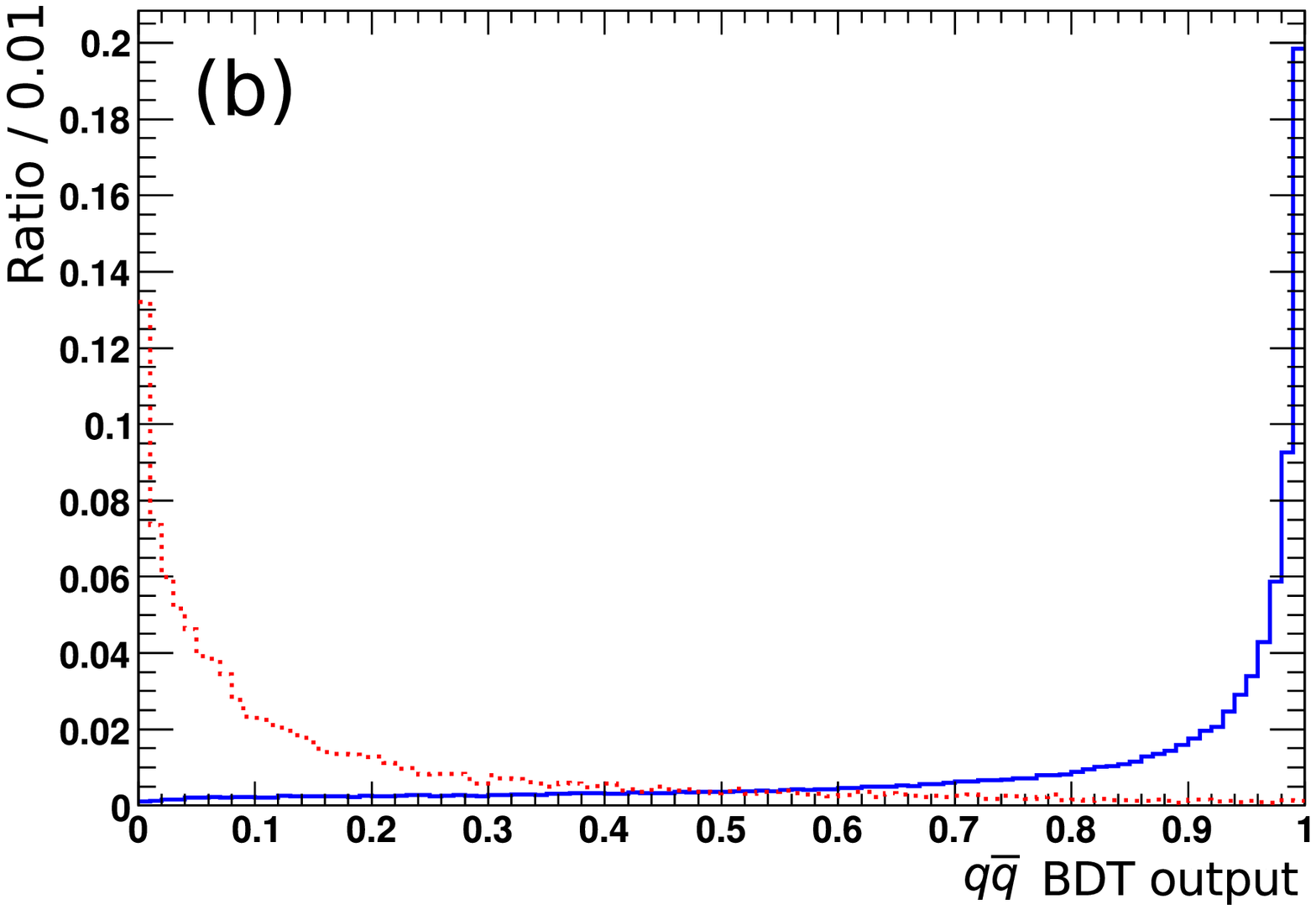}
\caption{
    The (a) $\BB$ and (b) $\qqbar$ $\epem$ BDT outputs for simulated events in the low $s$ region. Shown are the
    distributions for $\BB$ background (red dashed line),
    $\qqbar$ background (red dotted line), and signal (blue solid line) 
    event samples, normalized to unit area.}
\label{fig:bdtout}

\end{center}
\end{figure}

The main backgrounds arise from random combinations of leptons from semileptonic $B$ and $D$ decays.
These combinatorial backgrounds from either $\BB$ events
(referred to as ``$\BB$ backgrounds'') or continuum $\qqbar$ events ($\epem\to \qqbar,\ q=u,d,s,c$, referred to as ``$\qqbar$ backgrounds'') are suppressed using bagged decision trees (BDTs)~\cite{bdts}.
We train eight separate BDTs as follows:
\begin{itemize}
\item Suppression of $\BB$ backgrounds for $\epem$ modes in the low $s$ region;
\item Suppression of $\BB$ backgrounds for $\epem$ modes in the high $s$ region;
\item Suppression of $\BB$ backgrounds for $\mumu$ modes in the low $s$ region;
\item Suppression of $\BB$ backgrounds for $\mumu$ modes in the high $s$ region;
\item Suppression of $\qqbar$ backgrounds
for $\epem$ modes in the low $s$ region;
\item Suppression of $\qqbar$ backgrounds for $\epem$ modes in the high $s$ region;
\item Suppression of $\qqbar$ backgrounds for $\mumu$ modes in the low $s$ region;
\item Suppression of $\qqbar$ backgrounds for $\mumu$ modes in the high $s$ region.
\end{itemize}
\noindent  The BDT input parameters 
include the following observables:
\begin{itemize}
\item $\Delta E$ of the $\B$ candidate;
\item The ratio of Fox-Wolfram moments $R_2$~\cite{Foxwolfram}
and the ratio of the second-to-zeroth 
angular moments of the energy flow
$L_2/L_0$~\cite{LegMom},
both event shape parameters
calculated using charged and neutral particles in the CM frame;
\item The mass and $\Delta E$ of the other $B$ meson in the event (referred to as the ``rest of the event'') 
computed in the laboratory frame 
by summing the momenta and energies of all charged particles and photons that are not used to reconstruct the signal candidate;
\item The magnitude of the total transverse momentum of the event in the laboratory frame;
\item 
The probabilities that the $B$ candidate and the dilepton candidate, respectively, originate from a single point in space;
\item The cosine values of four angles: the angle between the $B$ candidate momentum and the beam axis, the angle between the event thrust axis and the beam axis, the angle between the thrust axis of the rest of the event and the beam axis,
    and the angle between the event thrust axis and the thrust axis of the rest of the event,
    all defined in the CM frame.
\end{itemize}
Figure~\ref{fig:bdtout} shows the output distributions of the BDTs 
for Monte Carlo (MC) simulated signal and combinatorial background for the $e^+ e^-$ sample below
the $\jpsi$ resonance.
The distributions are histograms normalized to unit area.
The selections on BDT outputs are further optimized to maximize
the statistical significance of the signal events,
    as shown later.

Another source of background arises from $B \to D(\to \Kmaybestar \pi) \pi$ decays if both
pions are misidentified as leptons. Determined from data control samples with high purity~\cite{BaBarDetector},
      the misidentification rates for muons and electrons are
$\sim 3\%$ and $\lesssim 0.1\%$ 
per candidate, respectively. Thus, this background is only significant for
 $\mu^+\mu^-$ final states. We veto these events by requiring the invariant mass of the
$\Kmaybestar\pi$ system to be outside the range $1.84-1.90$~\gevcc after assigning the pion mass hypothesis to the muon candidates. Any remaining residual backgrounds from this type of contribution
are parameterized using control samples obtained from data.

After applying all selection criteria about 85\% of signal
events contain more than one $B$ candidate. These candidates differ typically in one charged or neutral hadron.
The average number of candidates per signal event is about six.
To choose the best candidate, we define the ratio

\begin{equation}
{\lambda} \equiv \frac{{\cal P}_{sig}^{B \bar B} +{\cal P}_{sig}^{\qqbar }}{{\cal P}_{sig}^{B \bar B} +{\cal P}_{sig}^{\qqbar} +{\cal P}_{bkg}^{B \bar B} +{\cal P}_{bkg}^{\qqbar} },
\end{equation}

\noindent where ${\cal P}_{sig}$ and ${\cal P}_{bkg}$ are probabilities calculated from the corresponding $B \bar B$ and $\qqbar$ BDT output distributions for signal and background, respectively. 
We select the candidate with the largest $\lambda$ as the best candidate.
The probability for a correctly-reconstructed signal event to be selected as the best candidate
is mode-dependent and varies between about
$80\%$ and $95\%$ for $s$ bins below the $J/\psi$ mass, while for $s$ bins above the $\psitwos$ mass it varies between about $60\%$ and $90\%$.

\section{Selection Optimization}
\label{sec:cutopt}

To optimize the $\Delta E$ selection, we simultaneously vary the
upper and lower bounds of the $\Delta E$
interval to find the values that maximize
the ratio $S/\sqrt{S + B}$ in the signal region ($\mes > 5.27~{\rm GeV}/c^2$, and for $K^*$ modes in addition $0.78 < \mkpi <  0.97~{\rm GeV}/c^2$),
where $S$ and $B$ are the expected numbers~\cite{PDG} of signal and combinatorial background events, respectively. We perform separate optimizations for dilepton masses below and above the $J/\psi$ mass.
For some modes, the optimization tends to select very narrow intervals,
   which leads to small signal efficiency.  To prevent this, we require
   the magnitudes of the $\DeltaE$ upper and lower bounds to be 0.04~\gev or larger. (Note that the lower bound is always negative and
     the upper bound always positive.)

\begin{figure}
\begin{center}
\includegraphics[width=0.45\textwidth]{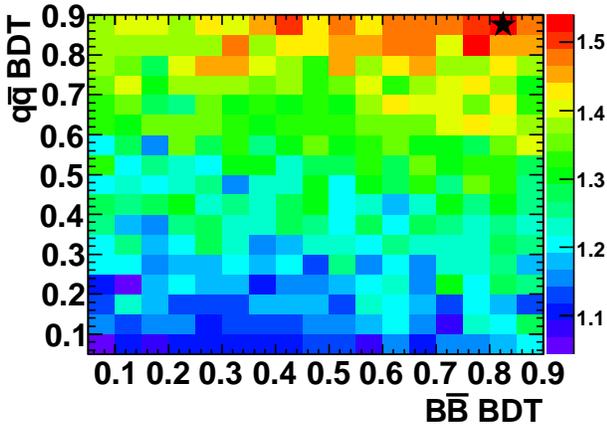}
\caption{Expected statistical significance of the number of fitted signal
events as a function of BDT interval lower bounds for $\modeeight$ in bin $s_2$. The star marks the
optimized pair of lower bounds.}
\label{fig:de-select}
\end{center}
\end{figure}

We also optimize the lower bounds on the BDT $\BB$ and $\qqbar$ intervals (the
   upper bounds on these intervals are always 1.0).  We perform fits to extract
   signal yields using the fit model described in Sec.~\ref{sec:fit}. For each mode, the
   lower bound on the BDT interval is optimized by maximizing the expected
   signal significance defined as the fitted signal yield divided by its associated uncertainty.
   We determine these from 500 pseudo-experiments using branching fraction averages~\cite{PDG}.
The optimized BDT lower bounds are listed in Tables~\ref{tab:kll-bdt} and \ref{tab:kstll-bdt} for $K \ell^+ \ell^-$ and
$K^* \ell^+ \ell^-$, respectively.
Figure~\ref{fig:de-select} shows the expected experimental significance in the $B \bar B$ BDT versus the $\qqbar$ BDT plane for  $\modeeight$ in bin $s_2$.
The signal selection efficiency and the cross-feed fraction 
        (defined in Sec.~\ref{sec:fit}) in each mode and $s$ bin after the final event selection are also
listed in Tables~\ref{tab:kll-bdt} and \ref{tab:kstll-bdt}. The selection efficiencies determined in simulations
vary from $11.4\pm 0.2$\% for $K^0_S \pi^+ e^+ e^-$ in $s_6$  to $33.3\pm0.3$\% for $K^+ \mumu$ in $s_5$, 
where the uncertainties are statistical.

\begin{table}
\centering
\caption{Optimized lower bounds
on the BDT intervals,
signal reconstruction efficiency, and
cross-feed fraction, 
by $K\ellell$ mode and $s$ bin. The uncertainties are statistical only.
}
\label{tab:kll-bdt}
\begin{tabular}{cccccc}
\\ \hline \hline \\
Mode & $s$ bin & $B \bar B$ & $\qqbar$& Efficiency& Cross-feed  \\
           &                &     BDT        &   BDT        &     $[\%]$  &     fraction $[\%]$ \\ \hline
$\modeone$
& $s_1$ & 0.20 & 0.80 & $19.9\pm 0.2$ & $8.9\pm 0.3$ \\
& $s_2$ & 0.70 & 0.85 & $22.2\pm 0.2$ & $8.6\pm 0.2$ \\
& $s_3$ & 0.20 & 0.85 & $25.2\pm 0.1$ & $8.9\pm 0.2$ \\
& $s_4$ & 0.70 & 0.70 & $24.3\pm 0.2$ & $9.4\pm 0.2$ \\
& $s_5$ & 0.70 & 0.80 & $22.2\pm 0.2$ & $12.0\pm 0.5$ \\
& $s_6$ & 0.75 & 0.80 & $16.6\pm 0.1$ & $21.7\pm 0.7$ \\
& $s_0$ & 0.50 & 0.85 & $22.7\pm 0.1$ & $8.8\pm 0.1$ \\\hline
$\modetwo$
& $s_1$ & 0.30 & 0.85 & $21.3\pm 0.2$ & $0.3\pm 0.0$ \\
& $s_2$ & 0.15 & 0.85 & $27.0\pm 0.2$ & $0.3\pm 0.0$ \\
& $s_3$ & 0.15 & 0.85 & $30.9\pm 0.1$ & $0.3\pm 0.0$ \\
& $s_4$ & 0.80 & 0.85 & $31.0\pm 0.2$ & $0.4\pm 0.0$ \\
& $s_5$ & 0.65 & 0.85 & $33.3\pm 0.3$ & $2.1\pm 0.1$ \\
& $s_6$ & 0.05 & 0.85 & $30.5\pm 0.2$ & $10.4\pm 0.2$ \\
& $s_0$ & 0.05 & 0.85 & $13.6\pm 0.1$ & $0.3\pm 0.0$ \\\hline
$\modethree$
& $s_1$ & 0.25 & 0.80 & $22.1\pm 0.2$ & $8.3\pm 0.3$ \\
& $s_2$ & 0.25 & 0.80 & $25.2\pm 0.2$ & $9.4\pm 0.3$ \\
& $s_3$ & 0.65 & 0.80 & $24.3\pm 0.1$ & $9.4\pm 0.2$ \\
& $s_4$ & 0.50 & 0.85 & $24.1\pm 0.2$ & $10.9\pm 0.4$ \\
& $s_5$ & 0.05 & 0.65 & $23.0\pm 0.2$ & $18.5\pm 0.9$ \\
& $s_6$ & 0.25 & 0.70 & $16.5\pm 0.1$ & $35.0\pm 1.1$ \\
& $s_0$ & 0.85 & 0.85 & $21.3\pm 0.1$ & $9.2\pm 0.2$ \\\hline
$\modefour$
& $s_1$ & 0.35 & 0.85 & $22.8\pm 0.2$ & $0.4\pm 0.1$ \\
& $s_2$ & 0.10 & 0.85 & $28.8\pm 0.2$ & $0.4\pm 0.0$ \\
& $s_3$ & 0.10 & 0.85 & $30.8\pm 0.1$ & $0.5\pm 0.0$ \\
& $s_4$ & 0.30 & 0.80 & $32.7\pm 0.2$ & $1.1\pm 0.1$ \\
& $s_5$ & 0.25 & 0.80 & $31.7\pm 0.3$ & $4.3\pm 0.2$ \\
& $s_6$ & 0.50 & 0.85 & $25.1\pm 0.2$ & $12.0\pm 0.3$ \\
& $s_0$ & 0.40 & 0.85 & $29.6\pm 0.1$ & $0.5\pm 0.0$ \\\hline          
\hline
\end{tabular}
\end{table}

\begin{table}
\centering
\caption{Optimized lower bounds
on the BDT intervals,
signal reconstruction efficiency, and
cross-feed fraction, 
by $K^*\ellell$ mode and $s$ bin. The uncertainties are statistical only.}
\label{tab:kstll-bdt}
\begin{tabular}{lccccc}
\\ \hline \hline \\
Mode & $s$ bin & $B \bar B$ &$\qqbar$ & Efficiency & Cross-feed \\
           &                &            BDT &    BDT       &     $[\%]$  &     fraction $[\%]$ \\ \hline
$\modeseven$
& $s_1$ & 0.55 & 0.85 & $13.6\pm 0.1$ & $14.0\pm 0.5$ \\
& $s_2$ & 0.80 & 0.85 & $14.6\pm 0.2$ & $19.2\pm 0.7$ \\
& $s_3$ & 0.85 & 0.80 & $14.9\pm 0.1$ & $20.7\pm 0.5$ \\
& $s_4$ & 0.85 & 0.85 & $14.7\pm 0.1$ & $28.0\pm 0.7$ \\
& $s_5$ & 0.15 & 0.85 & $16.4\pm 0.2$ & $59.3\pm 1.3$ \\
& $s_6$ & 0.10 & 0.85 & $14.3\pm 0.1$ & $110.8\pm 1.9$ \\
& $s_0$ & 0.80 & 0.85 & $14.5\pm 0.1$ & $18.9\pm 0.5$ \\\hline
$\modeeight$
& $s_1$ & 0.80 & 0.85 & $16.2\pm 0.1$ & $4.9\pm 0.2$ \\
& $s_2$ & 0.80 & 0.85 & $19.6\pm 0.2$ & $7.8\pm 0.3$ \\
& $s_3$ & 0.75 & 0.85 & $21.3\pm 0.1$ & $10.1\pm 0.2$ \\
& $s_4$ & 0.85 & 0.85 & $20.9\pm 0.1$ & $13.8\pm 0.3$ \\
& $s_5$ & 0.75 & 0.85 & $22.8\pm 0.2$ & $31.7\pm 0.6$ \\
& $s_6$ & 0.80 & 0.80 & $19.5\pm 0.2$ & $61.0\pm 0.9$ \\
& $s_0$ & 0.60 & 0.85 & $20.4\pm 0.1$ & $8.9\pm 0.2$ \\\hline
$\modeeleven$
& $s_1$ & 0.45 & 0.70 & $16.6\pm 0.2$ & $17.8\pm 0.6$ \\
& $s_2$ & 0.85 & 0.85 & $13.7\pm 0.2$ & $20.7\pm 0.8$ \\
& $s_3$ & 0.55 & 0.85 & $16.0\pm 0.1$ & $27.5\pm 0.7$ \\
& $s_4$ & 0.40 & 0.85 & $15.4\pm 0.1$ & $41.6\pm 0.9$ \\
& $s_5$ & 0.80 & 0.45 & $13.1\pm 0.2$ & $68.6\pm 1.8$ \\
& $s_6$ & 0.60 & 0.85 & $11.4\pm 0.2$ & $133.4\pm 2.9$ \\
& $s_0$ & 0.70 & 0.85 & $16.0\pm 0.1$ & $23.1\pm 0.5$ \\\hline
$\modetwelve$
& $s_1$ & 0.80 & 0.85 & $16.5\pm 0.2$ & $6.8\pm 0.2$ \\
& $s_2$ & 0.85 & 0.85 & $18.6\pm 0.2$ & $10.9\pm 0.3$ \\
& $s_3$ & 0.80 & 0.80 & $18.5\pm 0.1$ & $11.2\pm 0.3$ \\
& $s_4$ & 0.55 & 0.65 & $21.9\pm 0.2$ & $25.6\pm 0.4$ \\
& $s_5$ & 0.75 & 0.80 & $19.0\pm 0.2$ & $50.4\pm 0.9$ \\
& $s_6$ & 0.05 & 0.80 & $15.1\pm 0.2$ & $110.9\pm 1.8$ \\
& $s_0$ & 0.80 & 0.85 & $19.7\pm 0.1$ & $10.8\pm 0.2$ \\\hline           
\hline
\end{tabular}
\end{table}

\section{Fit Methodology}
\label{sec:fit}

We perform one-dimensional fits in $\mes$ for $K \ellell$ modes  and two-dimensional fits in  $\mes$  and $\mkpi$ for $K^* \ellell$ modes to extract the signal yields. The probability density function (PDF) for signal \mes is  parametrized by a Gaussian function with mean and width fixed to values obtained from fits to
the vetoed $J/\psi$ events in the data control samples.
For
$ \mkpi$, the PDF is a relativistic Breit-Wigner line shape~\cite{Aubert:2008gm}.
True signal events are those where all
generator-level final-state daughter particles are correctly reconstructed and are selected to form a $B$ candidate.

   For the combinatorial background, the $\mes$ PDF is modeled with a kinematic threshold function
whose shape is a free parameter in the fits~\cite{argus},
      while the  $\mkpi$ PDF shape is characterized with the $K^{*} e^\pm \mu^\mp $ sample mentioned in Sec.~\ref{sec:selection}.
We parameterize the combinatorial
$\mkpi$ distributions with non-parametric Gaussian kernel density estimator shapes~\cite{keys} (referred to as the ``KEYS PDFs'')
 drawn from the $K^{*} e^\pm \mu^\mp $ sample in the full $\mes$ fit region. Since the correlation between $\mkpi$ and $ \Delta E$ is weak,
 we accept all $K^{*} e^\pm \mu^\mp $ events within $|\Delta E|<0.3$~\gev,
 rather than imposing a stringent $\Delta E$ selection,
 in order to enhance sample sizes.

Signal cross-feed consists of mis-reconstructed signal events, in which typically a low-momentum $\pi^\pm$ or $\pi^0$ is swapped, added, or removed in the $B$ candidate reconstruction. We distinguish among
different categories of cross-feed: ``self-cross-feed'' is when a particle is swapped within one mode, ``feed-across'' is when a particle is swapped between two signal modes with the same final-state multiplicity, and ``feed-up (down)'' is when  a particle is added (removed) from a lower (higher) multiplicity $b \ra s \ellell$ mode.
We use both exclusive and inclusive $b \ra s \ellell$ MC samples to evaluate the contributions of the different categories.
The cross-feed $\mes$ distribution is typically broadened compared to correctly reconstructed signal decays.
We combine the cross-feed contributions from all sources into a single fit component that is modeled as a
sum of weighted histograms with a single overall normalization, which is allowed to scale as a fixed fraction of the observed correctly reconstructed signal yield. 
This fixed fraction is presented as the ``cross-feed fraction''
in Tables~\ref{tab:kll-bdt} and~\ref{tab:kstll-bdt}.
The modeling of cross-feed contributions is validated using fits to the vetoed $J/\psi K^{(*)}$ and $\psitwos K^{(*)}$ events, in which the cross-feed contributions are relatively large compared to all other backgrounds.

Exclusive $B$ hadronic decays may be mis-reconstructed as $B \ra K^{(*)}\ellell$, since hadrons can be misidentified as muons.
Following a procedure similar
to that described in Ref.~\cite{Aubert:2006vb}, we determine this background by selecting a sample of $K^{(*)}\mu^\pm h^\mp$ events,
in which the muon is identified as a muon and the hadron is inconsistent with an electron.
Requiring 
identified kaons and pions, we select subsamples of $K^{(*)}\pi^+ \pi^-,  K^{(*)}K^+ \pi^-, K^{(*)}\pi^+ K^-$, and $K^{(*)}K^+ K^-$.
We obtain weights from data control samples where a charged
particle's species can be identified with high precision and
accuracy without using particle identification information.
The weights are then applied to this dataset to characterize the contribution expected in our fits due to misidentified muon candidates.
We characterize the misidentification backgrounds 
using the KEYS PDFs,
   with normalizations obtained by construction directly from the weighted data.

Some charmonium events may escape the charmonium vetoes and appear in our fit region. Typically, this occurs when electrons radiate a photon or a muon candidate  is a misidentified hadron and the missing energy is accounted for by a low-energy $\pi^\pm$ or $\pi^0$. The largest background contributions from this source are expected
in the $K^*\mu^+\mu^-$ and $K^*e^+e^-$ channels. We model this background using the charmonium MC samples and determine the leakage into $s$ bins on either side of the $J/\psi$ and $\psitwos$ resonances. We see a
notable charmonium contribution (about five events) for $ B^0 \ra K^+\pi^-\mu^+\mu^-$ in bin $s_3$. 
This leakage is typically caused by a swap between the $\mup$ and $\pip$ in a single $B\to\jpsi(\to \mumu)K\pip$ candidate,
     where both the $\mup$ and $\pip$ are misidentified.

Hadronic peaking background from $B \ra K^* \pi^0$ and  $B \ra K^* \eta$ in which the $\pi^0$ or $\eta$ decays via Dalitz pairs shows a small peaking component in $\mes$ in bin $s_1$.
Due to the requirement $ s > 0.1~\rm GeV^2/c^4$, contributions of $\gamma$ conversions from $B  \ra K^*\gamma$ events beyond the photon pole region are found to be negligible.

\subsection*{Fit Model for Rate Asymmetries}

Using the PDFs described above, we perform simultaneous fits across different $K^{(*)} \ell^+\ell^-$ modes.
Since efficiency-corrected signal yields are shared across various decay modes, we can extract rate asymmetries
directly from the fits.
The fitted signal yields in $\Bp$ modes are corrected by the lifetime ratio $\tau_{\Bz}/\tau_{\Bp}$.
We also correct the signal yields for ${\cal B}(K^*\to K\pi)$ in $\Kstar$ modes
and ${\cal B}(\KS\to\pi^+\pi^-)$ in the modes with a $\KS$.
In the fits for ${\cal A}_{\CP}$, we share the efficiency-corrected signal yield $N_B$ as a floating variable for $B\ (q\bar{b}, q=u,d)$ events across different flavor-tagging $K^{(*)}\ell^+\ell^-$ modes by assuming lepton-flavor and isospin symmetries. The efficiency-corrected signal yield $N_{\bar B}$  for $\bar B\ (\bar{q} b)$ events is then defined by $N_{\bar B} = N_B \cdot (1 + {\cal A}_{\CP)})/(1 -{\cal A}_{\CP})$ and is also shared across corresponding
 modes. For the lepton-flavor ratios ${\cal R}_ {K^{(*)}}$, we share the efficiency-corrected signal yield $N_{ee}$ as a floating variable for the two $B  \ra K e^+e^-$ or $B \ra K^*e^+e^-$ modes by assuming isospin symmetry. The efficiency-corrected signal yield $N_{\mu \mu}$ shared across the corresponding $B \ra K^{(*)}\mu^+\mu^-$ modes is then defined by $N_{\mu \mu} = N_{ee} \cdot R_{ K^{(*)}}$. For the isospin asymmetry ${\cal A}^{K^{(*)}}_I$, we share the efficiency-corrected signal yield $N_{B^+}$ as a floating variable for the two $B^+  \ra K^+\ell^+\ell^-$ or $B^+ \ra K^{*+}\ell^+\ell^-$ modes by assuming lepton-flavor symmetry.
The efficiency-corrected signal yield $N_{B^0}$ shared across the corresponding $B^0  \ra K^{*0}\ell^+\ell^-$ modes is then defined by 
 $N_{B^0} = N_{B^+} \cdot (1 +{\cal A}^{K{(*)}}_I )/(1 -{\cal A}^{K{(*)}}_I )$.

\section{Fit Validation}
\label{sec:charmonium}

We  validate the fit methodology with charmonium control samples obtained from the dilepton mass regions around the $J/\psi$ and $\psitwos$ resonances that are vetoed in the $B \ra K^{(*)} \ell^+ \ell^-$ analysis.
We measure the $J/\psi K^{(*)}$ and $\psitwos K^{(*)}$ branching fractions in each final state
with the optimized BDT selections in bins $s_3$ and $s_4$, respectively. 
Our measurements agree well with the world averages~\cite{PDG} for all final states. Typical deviations, based on statistical uncertainties only, are less than one standard deviation ($\sigma$). 
The largest deviation, in the $K^+ \pi^- \mu^+ \mu^-$ mode, is $1.7\sigma$. For $J/\psi K^{(*)}$ modes, the statistical uncertainties are considerably smaller than those of the world averages.
We float the Gaussian means and widths of the signal PDFs 
in the fits for the $J/\psi K^{(*)}$ modes.
The associated uncertainties obtained from the fits are then used as a source of systematic variation for the signal PDFs. The typical signal width in $\mes$ is $2.5 ~{\rm MeV}/c^2$.

We further validate our fitting procedure by applying it to charmonium events to extract the rate asymmetries. The measured \CP asymmetries ${\cal A}_{CP}$, lepton-flavor ratios ${\cal R}_{K^{(*)}}$ and isospin asymmetries ${\cal A}_I$ are in good agreement
with Standard Model expectations or world averages for~${\cal A}_I$.

We also test the methodology with fits to ensembles of datasets where signal and background events are generated from appropriately normalized PDFs (``pure pseudo-experiments''). We perform fits to these
pseudo-experiments in each mode and $s$ bin using the full fit model described previously. 
For ensembles of 1000 pure pseudo-experiments, the pull distributions for the signal yields show negligible biases.
We further fit ensembles of pseudo-experiments in which the signal events are drawn from properly normalized exclusive MC samples (``embedded pseudo-experiments''). The pull distributions also show the expected performance.

We perform fits to ensembles of pure pseudo-experiments in order to estimate the statistical sensitivity of,
and biases related to, the various rate asymmetry measurements. 
The pull distributions for ${\cal A}_{\CP}$ and ${\cal R}_{K^{(*)}}$ for 
the low and high $s$ regions show minimal biases.
For ${\cal A}_I$, we test a series of ${\cal A}_I$  input values ($-0.6,\ -0.3,\ 0.0,\ 0.3,\ 0.6$) in each $s$ bin
using pure pseudo-experiments to ensure we obtain unbiased fits under different assumptions of isospin asymmetry. The ${\cal A}^K_I$ pulls
are generally well-behaved.
In the worst case, the test fits for ${\cal A}^K_I$ are slightly biased due to
very low signal yield expectations in the $\KS\ellell$ final states.

\section{Systematic Uncertainties}
\label{sec:systematic}

Since some systematic uncertainties largely cancel in ratios, it is useful to separate
the discussion of systematic uncertainties on partial branching fractions
from that on rate asymmetries.

\subsection{Branching Fraction Uncertainties}
\label{sec:bfsystematic}

Systematic uncertainties for branching fractions arise from multiplicative
systematic uncertainties involving the determination of the signal efficiency,
and from additive systematic uncertainties arising from the extraction of signal
yields in the data fits.
The multiplicative systematic errors include contributions from the

\begin{itemize}
\item Number of \BB pairs: This uncertainty is $0.6\%$.

\item Tracking efficiency for charged particles:
We assign a correlated uncertainty of 0.3\%
for each lepton, and 0.4\% for each charged hadron
    including daughter pions from \KS decay~\cite{Allmendinger:2012ch}.

\item Charged particle identification (PID) efficiencies:
We employ a data-driven method to correct PID efficiencies in simulated events. We estimate the systematic uncertainties from the change in signal efficiency for  simulated $\jpsi\Kmaybestar$ events after turning off the PID corrections.
The systematic uncertainties are mode dependent and vary between $0.3\%$ and $1.6\%$.

\item \KS identification efficiency: This is determined as a
function of flight distance after applying \KS efficiency corrections.
An uncertainty of $0.9\%$ is obtained by varying the \KS selection algorithm.

\item Event selection efficiency:
We measure the efficiency of the BDT selection in charmonium data control
samples and compare with results obtained
for exclusive charmonium samples from simulation. We take the
magnitude of the deviation 
for any particular final state
and $s$ bin as the uncertainty associated with the BDT lower bounds.
If the data and simulation are consistent within the uncertainty,
we then take the uncertainty as the systematic uncertainty.
The systematic uncertainty is found to vary between 0.3\% and 9.1\%
depending on both the mode and the $s$ bin. Due to a strong correlation
between the $\DeltaE$ and BDT outputs, uncertainties due to $\Delta E$ are fully accounted for by this procedure.

\item Monte Carlo sample size:
We find the uncertainty related to the
   finite size of the MC sample to be of the order of 1\% or less
for all modes.

\end{itemize}

The additive systematic uncertainties involve contributions from the

\begin{itemize}

\item Signal PDF shapes:
We characterize them by varying the PDF shape parameters (signal mean, signal width, and combinatorial background shape and normalization) by the statistical uncertainties obtained in the fits to
the \jpsi data control samples for \mes and signal MC events for \mkpi.

\item Hadronic backgrounds: We characterize them
by varying both the normalization by the associated statistical
uncertainties
and by performing fits with different choices of smoothing parameters
for the KEYS PDF shapes.

\item Peaking backgrounds from charmonium events and $\piz/\eta$ Dalitz decays: We vary the normalization
for these contributions by $\pm$25\%.

\item Modeling of \mkpi line shapes of the
combinatorial background: We characterize the uncertainties by analyzing
data samples 
selected from the $\mes <5.27$~\gevcc sideband,
     and simulated events.

\end{itemize}

Table~\ref{tab:bfsystm} summarizes all sources of systematic uncertainties
considered in the total branching fraction measurements for individual modes.
The total systematic uncertainty for the branching fractions is 
obtained by summing in quadrature
the above-described uncertainties from different categories.

\begin{table*}
\caption{Individual systematic uncertainties [\%] for measurements of the total branching fractions in $\Kmaybestar\ellell$ decays.}
\centering\footnotesize
\begin{tabular}{lrrrrrrrr}
\hline \hline
Mode & $\modeoneshort$ & $\modetwoshort$ & $\modethreeshort$ & $\modefourshort$ & $\modesevenshort$ & $\modeeightshort$ & $\modeelevenshort$ & $\modetwelveshort$\\ \hline
$\BB$ counting & $\pm0.6$ & $\pm0.6$ & $\pm0.6$ & $\pm0.6$ & $\pm0.6$ & $\pm0.6$ & $\pm0.6$ & $\pm0.6$ \\
Tracking & $\pm1.4$ & $\pm1.0$ & $\pm1.4$ & $\pm1.0$ & $\pm1.8$ & $\pm1.4$ & $\pm1.8$ & $\pm1.4$ \\
PID & $\pm1.6$ & $\pm0.3$ & $\pm0.7$ & $\pm0.4$ & $\pm1.5$ & $\pm0.3$ & $\pm0.5$ & $\pm1.2$ \\
$K^0_s$ ID & $\pm0.9$ & --- & $\pm0.9$ & --- & $\pm0.9$ & --- & $\pm0.9$ & --- \\
BDT selections & $\pm2.2$ & $\pm1.7$ & $\pm4.7$ & $\pm1.5$ & $\pm8.3$ & $\pm2.5$ & $\pm9.1$ & $\pm2.7$ \\
MC sample size & $\pm0.3$ & $\pm0.3$ & $\pm0.3$ & $\pm0.3$ & $\pm0.4$ & $\pm0.3$ & $\pm0.4$ & $\pm0.4$ \\
\hline
Sig. Shape & $\pm0.5$ & $\pm0.4$ & $\pm1.5$ & $\pm0.4$ & $\pm1.5$ & $\pm0.7$ & $\pm1.5$ & $\pm0.7$ \\
Hadronic & $\pm3.3$ & $\pm5.8$ & --- & --- & $\pm2.3$ & $\pm1.6$ & --- & --- \\
Peaking & $\pm0.3$ & $\pm0.8$ & $\pm1.2$ & $\pm0.8$ & $\pm0.7$ & $\pm1.7$ & $\pm0.8$ & $\pm1.2$ \\
Comb. $\mkpi$ shape & --- & --- & --- & --- & $\pm1.2$ & $\pm0.6$ & $\pm0.6$ & $\pm1.6$ \\
\hline
Total & $\pm4.7$ & $\pm6.3$ & $\pm5.4$ & $\pm2.2$ & $\pm9.3$ & $\pm3.9$ & $\pm9.5$ & $\pm4.0$ \\
\hline \hline
\end{tabular}
\label{tab:bfsystm}
\end{table*}

\subsection{Systematic uncertainties for the rate asymmetries}

For ${\cal A}_{\CP}$, a large portion of the uncertainties associated with
the signal efficiency cancel.  We find that the only efficiency-related term discussed in Sec.~\ref{sec:bfsystematic}
 that is not negligible for ${\cal A}_{\CP}$ is 
 the one associated with the PID selection.  Amongst the efficiency-related systematics, we therefore only consider this term.  We also
 consider the additive systematic uncertainties listed in 
 Sec.~\ref{sec:bfsystematic}.
 Our measured ${\cal A}_{\CP}$ central
 values for $\jpsi K$ and $\jpsi K^*$ are both well below 1\%
 and show minimal detector efficiency effects.
Potential, additional ${\cal A}_{\CP}$ systematic effects from the assumptions of lepton-flavor and isospin symmetry are tested by removing these
assumptions.

The systematic uncertainties for the
lepton-flavor ratios ${\cal R}_{K^{(*)}}$ are calculated by summing
in quadrature the systematic errors in the
muon and electron modes. Common systematic effects,
such as tracking, $\KS$ efficiency, and $\BB$ counting,
    yield negligible uncertainties in the ratios.
    Potential, additional ${\cal R}_{K^{(*)}}$ systematic effects are tested
by removing the assumption of isospin symmetry.

For the systematic uncertainties of ${\cal A}_I$, we sum
in quadrature the systematic errors in charged and neutral $B$ modes. Common systematic effects,
which include $\BB$ counting and a large portion of the uncertainties 
associated with
PID and tracking efficiencies, are negligible. Again, additional tests on ${\cal A}_I$ systematics
are performed by relaxing the assumption of lepton-flavor symmetry. 
Furthermore, as the cross-feed fractions in Tables~\ref{tab:kll-bdt} and~\ref{tab:kstll-bdt} 
are estimated under the assumption of isospin symmetry, we test this systematic effect using cross-feed fractions estimated with different 
${\cal A}_I$ input values.

Our checks on symmetry assumptions described above
for ${\cal A}_{\CP}$, ${\cal R}_{\Kmaybestar}$ and ${\cal A}_I$ generally show deviations
from the original measured values below 20\% of the associated statistical
uncertainties, 
and so we do not assign additional uncertainties.

\section{Results}
\label{sec:result}

We perform fits for each $\Kmaybestar\ellell$ final state in each $s$ bin listed
in Tables~\ref{tab:kll-bdt} and~\ref{tab:kstll-bdt} 
to obtain signal and background yields, $N_{\rm sig}$ and $N_{\rm bkg}$, respectively. We model the different background components by the PDFs described
in Sec.~\ref{sec:fit}.
We allow the shape parameter of the $\mes$ kinematic threshold function of the combinatorial background to float in the fits. For the signal, we use a fixed Gaussian shape unique to each final state, as described previously.
We leave the shapes
of the other background PDFs fixed. For the peaking background, we fix the
absolute normalization. For the cross-feed, we fix the normalization relative to the signal yields.

\begin{figure}[b!]
\begin{center}
\includegraphics[width=0.45\textwidth]{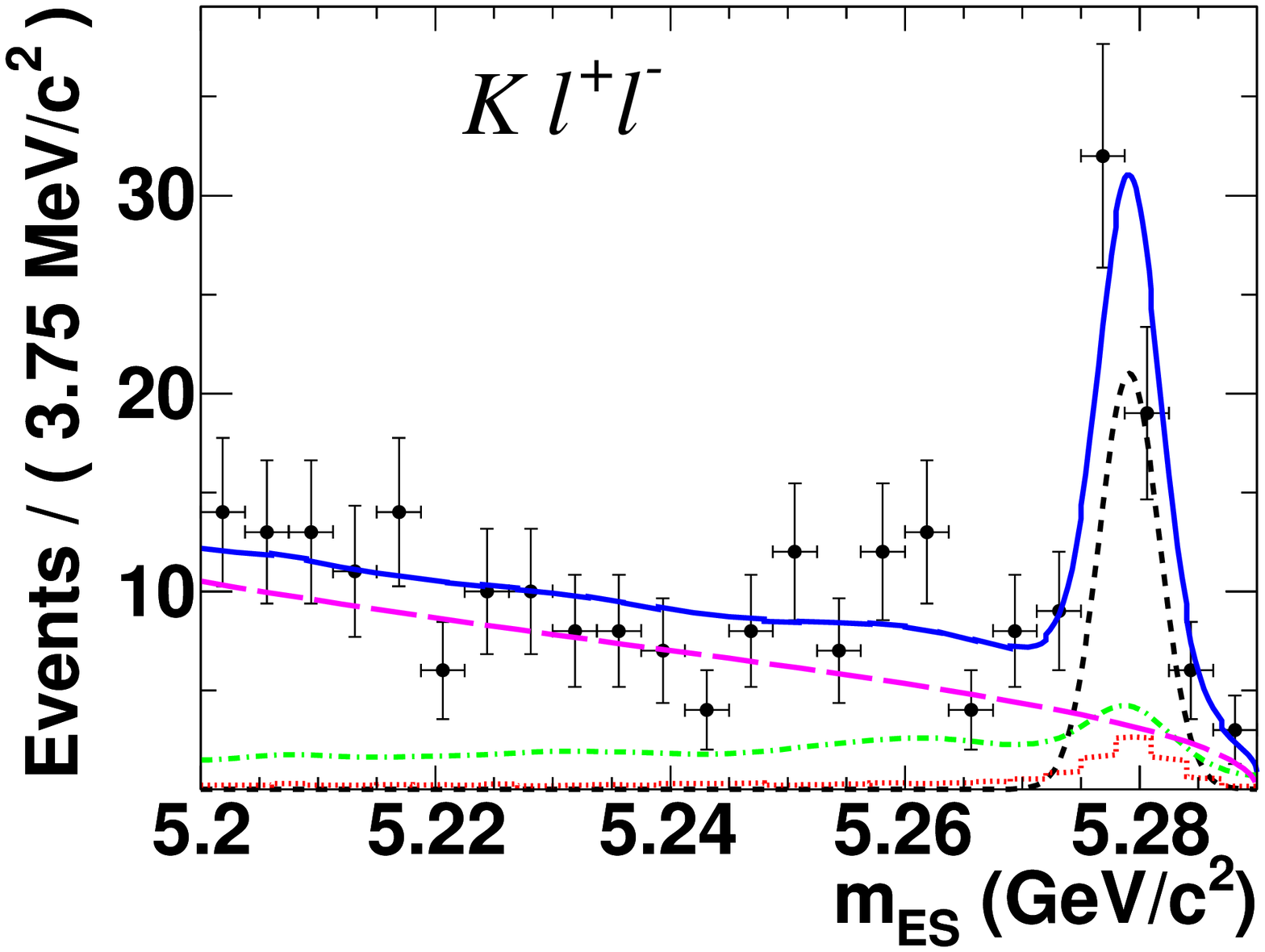}
\caption{The $\mes$ spectrum in bin $s_4$ for all $K \ell^+ \ell^-$ modes combined showing data (points with error bars), the total fit (blue solid line), signal component (black short-dashed line), combinatorial background (magenta long-dashed line), hadrons misidentified as muons (green dash-dotted line), and the sum of cross-feed and peaking components (red dotted line). }
\label{fig:kll4}
\end{center}
\end{figure}

\begin{figure}[b!]
\begin{center}
\includegraphics[width=0.5\textwidth]{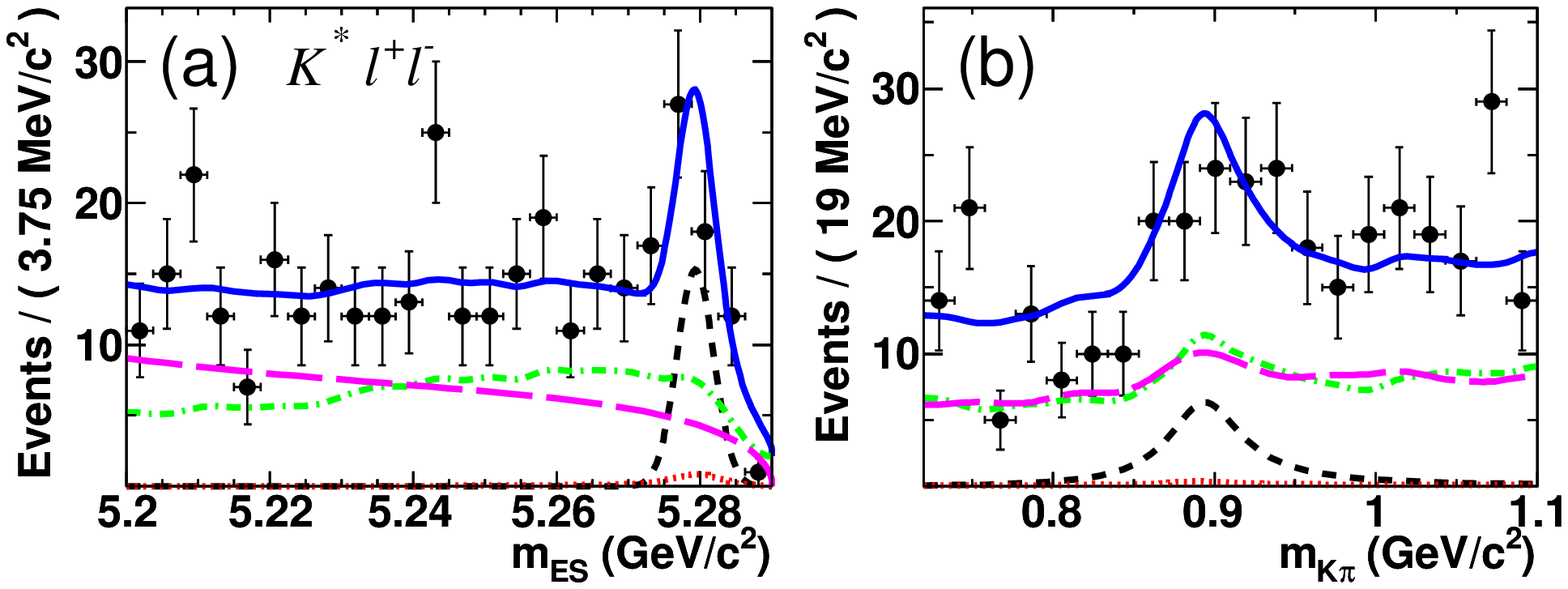}
\caption{The (a) $\mes$ and (b) $\mkpi$ mass spectra in bin $s_1$ for all four $K^* \ell^+ \ell^-$ modes combined showing data (points with error bars), the total fit (blue solid lines), signal component (black short-dashed lines), combinatorial background (magenta long-dashed lines), hadrons misidentified as muons (green dash-dotted lines), and the sum of cross-feed and peaking components (red dotted lines). }
\label{fig:kstll1}
\end{center}
\end{figure}

\begin{figure}[b!]
\begin{center}
\includegraphics[height=8.0cm,width=0.45\textwidth]{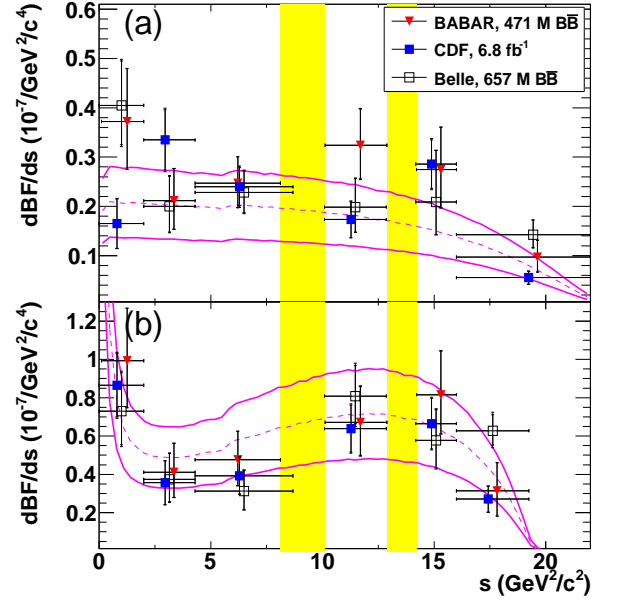}
\caption{Partial branching fractions for the (a) $K \ell^+ \ell^-$ and (b) $K^* \ell^+ \ell^-$ modes as a function of $s$ showing \babar\, measurements (red triangles), Belle measurements~\cite{belle09} (open squares), CDF measurements~\cite{cdf10} (blue solid squares), and the SM prediction from the Ali~\etal~model~\cite{Ali:2002jg} with $B\to\Kmaybestar$ form factors~\cite{ffmodels} (magenta dashed lines).
       The magenta solid lines show the theory uncertainties. The vertical 
       yellow shaded bands show the vetoed $s$ regions around the $J/\psi$ and $\psitwos$. }
\label{fig:kllpbf}
\end{center}
\end{figure}

Figure~\ref{fig:kll4} shows as an example the $\mes$ distribution for the combined $K \ell^+ \ell^-$ modes in bin $s_4$, while Fig.~\ref{fig:kstll1} shows the $\mes$ and $\mkpi$ mass spectra for the combined $K^* \ell^+ \ell^-$ modes in bin $s_1$.
The cross-feed contributions and the peaking backgrounds are negligible
for this fit. The combinatorial background dominates and for $\mu^+ \mu^-$ modes misidentified hadrons are the second largest background.
From the yields in each $s$ bin we determine the partial branching fractions summarized in Table~\ref{tab:pbf}.
Figure~\ref{fig:kllpbf} shows our results for the partial branching fractions of the $K \ell^+ \ell^-$ and $K^* \ell^+ \ell^-$ modes in comparison to results from the Belle and CDF Collaborations~\cite{belle09, cdf10}
and to the prediction of the Ali~\etal~model~\cite{Ali:2002jg}. 
Our results are seen to agree with those of Belle and CDF.
Our results are also
in agreement with the most recent partial branching fraction measurements of
$B^0 \to K^{*0}\mumu$ from LHCb~\cite{lhcb11}.

\begin{figure}[b!]
\begin{center}
\includegraphics[height=4.0cm]{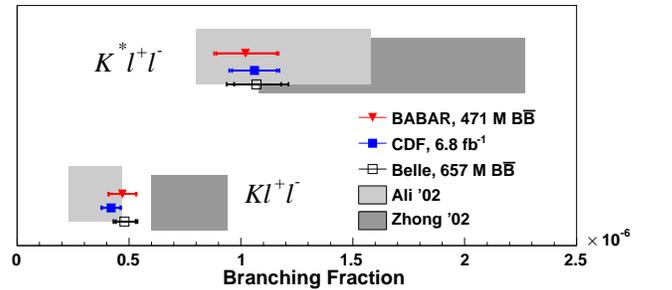}
\caption{Total branching fractions for the 
    $K \ell^+ \ell^-$ and  $K^* \ell^+ \ell^-$ modes (red triangles) compared
       with Belle~\cite{belle09} (open squares) and CDF~\cite{cdf10} (blue solid squares)
       measurements and with 
    predictions from the Ali~\etal~\cite{Ali:2002jg} (light grey bands), and Zhong~\etal~\cite{Zhong:2002nu} (dark grey bands) models. }
\label{fig:klltbf}
\end{center}
\end{figure}

The total branching fractions are measured to be
\begin{eqnarray}
{\cal B} (\modekavgll) & = & (4.7\pm0.6\pm 0.2) \times 10^{-7},\nonumber\\
{\cal B} (\modekstll) & = & (10.2_{-1.3}^{+1.4}\pm 0.5) \times 10^{-7}.\nonumber
\end{eqnarray}
Here, the first uncertainties are statistical, and the second are systematic.
The total branching fractions are shown in Fig.~\ref{fig:klltbf} in comparison to measurements from Belle~\cite{belle09} and CDF~\cite{cdf10}
and predictions from Ali~\etal~\cite{Ali:2002jg} and Zhong~\etal~\cite{Zhong:2002nu}.

\renewcommand\arraystretch{1.3}

\begin{table}[b!]
\footnotesize
\centering
\caption{Measured branching fractions [$10^{-7}$] by mode and $s$ bin. The first and second uncertainties are statistical and systematic, respectively.}
\label{tab:pbf}
\begin{tabular}{l|cc|cc} \hline \hline
& \multicolumn{2}{c}{$\modekavgll$} & \multicolumn{2}{|c}{$\modekstll$}\\
$s$ (\gevcccc) & $N_{\rm sig}$ & ${\cal B} [10^{-7}]$ & $N_{\rm sig}$ & ${\cal B} [10^{-7}]$ \\ \hline
0.10--2.00 & $20.6_{-5.4}^{+5.9}$ & $0.71_{-0.18}^{+0.20}\pm 0.02$ & $26.0_{-6.4}^{+7.1}$ & $1.89_{-0.46}^{+0.52}\pm 0.06$  \\
2.00--4.30 & $17.4_{-4.8}^{+5.4}$ & $0.49_{-0.13}^{+0.15}\pm 0.01$ & $14.5_{-4.6}^{+5.3}$ & $0.95_{-0.30}^{+0.35}\pm 0.04$  \\
4.30--8.12 & $37.1_{-7.5}^{+8.0}$ & $0.94_{-0.19}^{+0.20}\pm 0.02$ & $29.3_{-8.3}^{+9.1}$ & $1.82_{-0.52}^{+0.56}\pm 0.09$  \\
10.11--12.89 & $36.0_{-7.6}^{+8.2}$ & $0.90_{-0.19}^{+0.20}\pm 0.04$  & $31.6_{-8.1}^{+8.8}$ & $1.86_{-0.48}^{+0.52}\pm 0.10$ \\
14.21--16.00 & $19.7_{-5.6}^{+6.2}$ & $0.49_{-0.14}^{+0.15}\pm 0.02$  & $24.1_{-6.0}^{+6.7}$ & $1.46_{-0.36}^{+0.41}\pm 0.06$  \\
$>$16.00 & $22.3_{-6.9}^{+7.7}$ & $0.67_{-0.21}^{+0.23}\pm 0.05$  & $14.1_{-5.9}^{+6.6}$ & $1.02_{-0.42}^{+0.47}\pm 0.06$  \\\hline
1.00--6.00 & $39.4_{-7.1}^{+7.7}$ & $1.36_{-0.24}^{+0.27}\pm 0.03$ & $33.1_{-7.8}^{+8.6}$ & $2.05_{-0.48}^{+0.53}\pm 0.07$  \\
\hline \hline
\end{tabular}
\end{table}

\begin{figure}[b!]
\begin{center}
\includegraphics[width=0.5\textwidth]{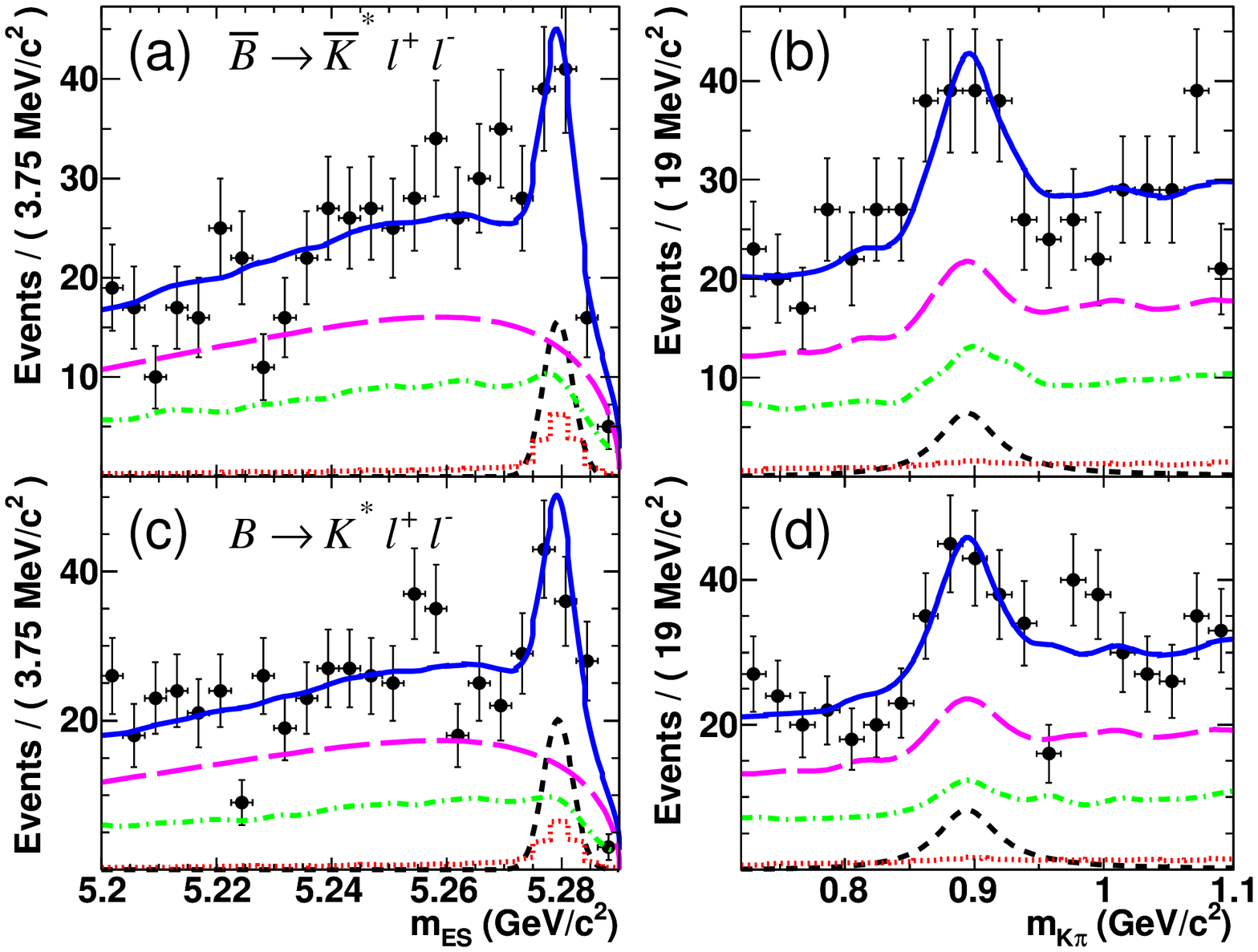}
\caption{(a)\&(c) \mes and (b)\&(d) \mkpi fits for ${\cal A}_{\CP}$ in the 
    (a)\&(b) $\bar{B}$ and (c)\&(d) $B$ low $s$ region for all four
    $K^* \ell^+ \ell^-$ modes combined. Data (points with error bars) are shown together with total fit (blue solid lines),
combinatorial background (magenta long-dashed lines),
signal (black short-dashed lines),
hadronic background (green dash-dotted lines),
and the sum of cross-feed and peaking background (red dotted lines). }
\label{fig:acpkstlllow}
\end{center}
\end{figure}

\begin{figure}[b!]
\begin{center}
\includegraphics[height=8cm,width=0.45\textwidth]{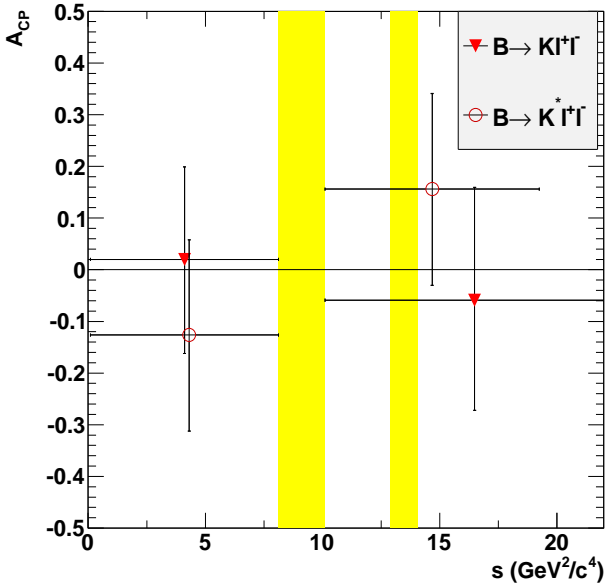}
\caption{\CP asymmetries ${\cal A}_{CP}$ for $K \ell^+ \ell^-$ modes (red solid triangles) and  $K^* \ell^+ \ell^-$ modes (red open circles) as a function of $s$. The vertical 
       yellow shaded bands show the vetoed $s$ regions around the $J/\psi$ and $\psitwos$.  }
\label{fig:acp}
\end{center}
\end{figure}

To measure direct ${\cal A}_{\CP}$, we fit the $B$ and $\bar B $ samples in the two $K^+\ellell$ modes and
four $\Kstar\ellell$ modes listed in Sec.~\ref{sec:selection}.
We perform the measurements in the full $s$ region,
   as well as in the low $s$
    and high $s$ regions
    separately.
The $B$ and $\bar B$ data sets share the same background shape parameter for the kinematic threshold function.
Figure~\ref{fig:acpkstlllow} shows an example fit for the
combined $\modekstll$ modes in the low $s$
    region.
Table \ref{tab:acp} summarizes the results. Figure~\ref{fig:acp} shows ${\cal A}_{\CP}$ as a function of $s$. Our results are consistent with the SM
expectation of negligible direct ${\cal A}_{\CP}$.

\renewcommand\arraystretch{1.2}
\begin{table}[b]
\centering
\caption{Measured ${\cal A}_{\CP}$ by mode and $s$ region. The first and second uncertainties are statistical and systematic, respectively. ``All'' refers to the union of $0.10<s<8.12$~\gevcccc and $s>10.11$~\gevcccc.}
\label{tab:acp}
\begin{tabular}{l|c|c} \hline \hline
$s$ (\gevcccc) & ${\cal }A_{\CP}(B^+ \to K^+ \ellell)$ & $A_{\CP}(\modekstll)$ \\ \hline
All & $-0.03\pm 0.14\pm 0.01$ & $0.03\pm 0.13 \pm 0.01$ \\ \hline
0.10--8.12 & $0.02\pm 0.18 \pm 0.01$ & $-0.13_{-0.19}^{+0.18}\pm 0.01$ \\
$>$10.11 & $-0.06_{-0.21}^{+0.22}\pm 0.01$ & $0.16_{-0.19}^{+0.18}\pm 0.01$  \\
\hline \hline
\end{tabular}
\end{table}

\begin{figure}[b!]
\begin{center}
\includegraphics[width=0.5\textwidth]{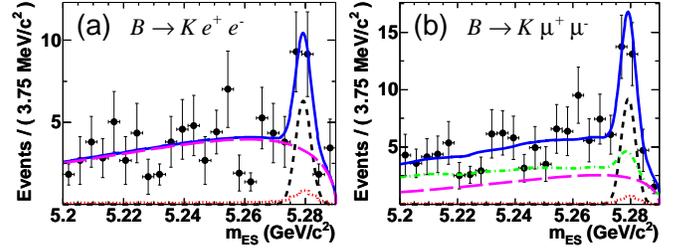}
\caption{\mes fits for ${\cal R}_K$ in the (a) $K\epem$ and
    (b) $K\mumu$ modes
    in the high $s$ region. Data (points with error bars) are shown together with total fit (blue solid lines),
combinatorial background (magenta long-dashed lines),
signal (black short-dashed lines),
hadronic background (green dash-dotted lines),
and the sum of cross-feed and peaking background (red dotted lines). }
\label{fig:emukllhi}
\end{center}
\end{figure}

\begin{figure}[b!]
\begin{center}
\includegraphics[height=8cm,width=0.45\textwidth]{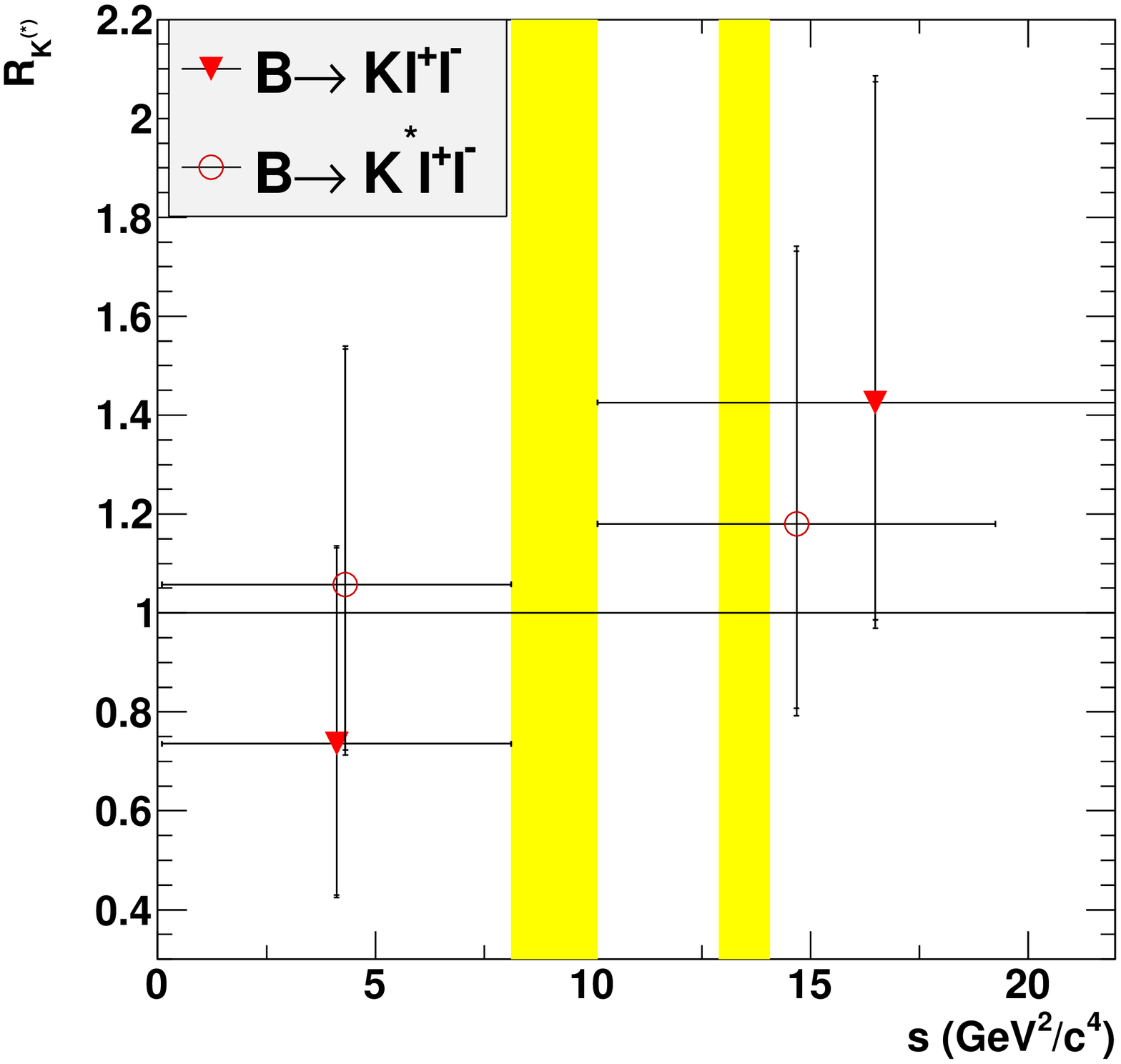}
\caption{Lepton flavor ratios ${\cal R}_{K^{(*)}}$ for the $K \ell^+ \ell^-$ (red solid triangles) and  $K^* \ell^+ \ell^-$ modes (red open circles) as a function of $s$. The vertical 
       yellow shaded bands show the vetoed $s$ regions around the $J/\psi$ and $\psitwos$. }
\label{fig:R_k}
\end{center}
\end{figure}

We fit the $e^+e^-$ and $\mu^+ \mu^-$ samples in the four $K \ell^+ \ell^-$ modes
and four $K^* \ell^+ \ell^-$ modes in the low $s$
    and high $s$ regions
    separately to measure the lepton-flavor ratios.
Figure~\ref{fig:emukllhi} shows an example fit for the
combined $K\mumu$ and $K\epem$ modes in the high $s$
    region.
Table~\ref{tab:emu} and Fig.~\ref{fig:R_k} show ${\cal R}_{K}$ and ${\cal R}_{K^*}$ for $s>0.1 \gevcccc$.
Our results are consistent with unity as expected in the SM.

\begin{table}[b!]
\centering
\caption{Measured ${\cal R}_{\Kmaybestar}$ by mode and $s$ region. The first and second uncertainties are statistical and systematic, respectively.  ``All'' refers to the union of $0.10<s<8.12$~\gevcccc and $s>10.11$~\gevcccc.}
\label{tab:emu}
\begin{tabular}{l|c|c} \hline \hline
$s$ (\gevcccc) & ${\cal R}_K$ & $R_{K^*}$\\ \hline
All & $1.00_{-0.25}^{+0.31}\pm 0.07$ &  $1.13_{-0.26}^{+0.34}\pm 0.10$ \\ \hline
0.10--8.12 &  $0.74_{-0.31}^{+0.40}\pm 0.06$ & $1.06_{-0.33}^{+0.48}\pm 0.08$ \\
$>$10.11 & $1.43_{-0.44}^{+0.65}\pm 0.12$ & $1.18_{-0.37}^{+0.55}\pm 0.11$ \\
\hline \hline
\end{tabular}
\end{table}

\begin{figure}[b!]
\begin{center}
\includegraphics[width=0.5\textwidth]{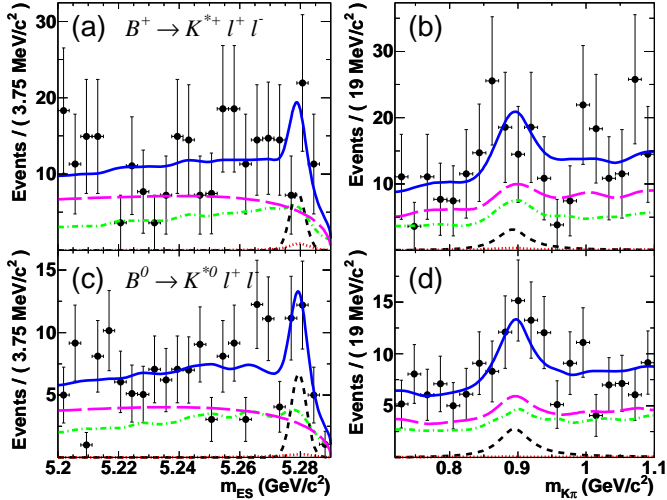}
\caption{The \mes and \mkpi fit projections for the
    (a)\&(b) $\Kstarp\ellell$
    and (c)\&(d) $\Kstarz\ellell$ modes in bin $s_2$.
        Data (points with error bars) are shown together with total fit (blue solid lines),
combinatorial background (magenta long-dashed lines),
signal (black short-dashed lines),
hadronic background (green dash-dotted lines),
and the sum of cross-feed and peaking background (red dotted lines). }
\label{fig:isokstll2}
\end{center}
\end{figure}

\begin{figure}[b!]
\begin{center}
\includegraphics[height=8cm,width=0.45\textwidth]{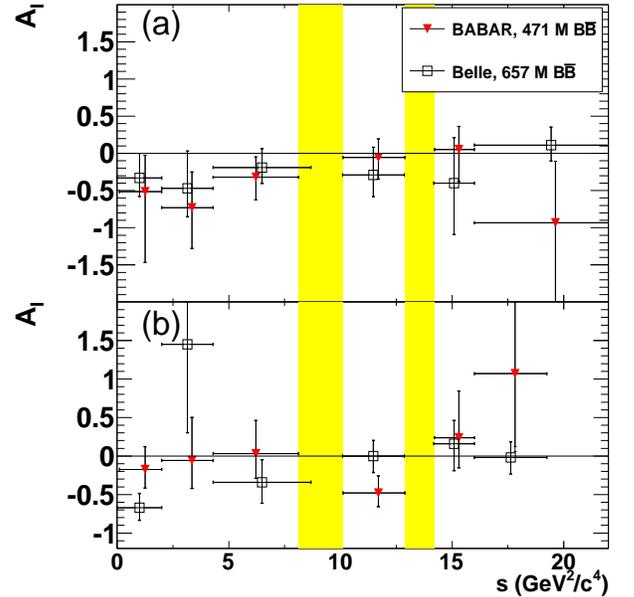}
\caption{Isospin asymmetry ${\cal A}_I$ for the (a) $K \ell^+ \ell^-$ and (b) $K^* \ell^+ \ell^-$ modes as a function of $s$ (red triangles), 
    in comparison to results from Belle~\cite{belle09} (open squares). The vertical 
       yellow shaded bands show the vetoed $s$ regions around the $J/\psi$ and $\psitwos$. }
\label{fig:isospin}
\end{center}
\end{figure}

We fit the data in each $s$ bin separately to determine ${\cal A}_I$ for
the
four combined $K \ell^+ \ell^-$ and four combined $K^* \ell^+ \ell^-$ modes.
Figure~\ref{fig:isokstll2}
shows an example fit for 
bin $s_2$.
The results are summarized in Table~\ref{tab:isospin}. Figure~\ref{fig:isospin} shows our measurements as a function of $s$
in comparison
with those of Belle~\cite{belle09}.
The two sets of results are seen to agree within the uncertainties.
Our results are also consistent with the SM prediction
that ${\cal A}_I$ is slightly negative ($\sim -1\%$) except in bin $s_1$,
where it is predicted to have a value around $+5\%$~\cite{isospin}.  

\begin{table}[b]
\centering
\caption{Measured ${\cal A}_I$ by mode and $s$ bin. The first and second uncertainties are statistical and systematic, respectively. }
\label{tab:isospin}
\begin{tabular}{l|cc} \hline \hline
 & \multicolumn{2}{c}{${\cal A}_I$}  \\
$s$ (\gevcccc) & $\modekavgll$ & $\modekstll$\\ \hline
0.10--2.00 & $-0.51_{-0.95}^{+0.49}\pm 0.04$ & $-0.17_{-0.24}^{+0.29}\pm 0.03$ \\
2.00--4.30 & $-0.73_{-0.55}^{+0.48}\pm 0.03$ & $-0.06_{-0.36}^{+0.56}\pm 0.05$ \\
4.30--8.12 & $-0.32_{-0.30}^{+0.27}\pm 0.01$ & $0.03_{-0.32}^{+0.43}\pm 0.04$ \\
10.11--12.89 & $-0.05_{-0.29}^{+0.25}\pm 0.03$ & $-0.48_{-0.18}^{+0.22}\pm 0.05$ \\
14.21--16.00 & $0.05_{-0.43}^{+0.31}\pm 0.03$ & $0.24_{-0.39}^{+0.61}\pm 0.04$ \\
$>$16.00 & $-0.93_{-4.99}^{+0.83}\pm 0.04$ & $1.07_{-0.95}^{+4.27}\pm 0.35$ \\\hline
1.00--6.00 & $-0.41\pm 0.25 \pm 0.01$ & $-0.20_{-0.23}^{+0.30}\pm 0.03$ \\
\hline \hline
\end{tabular}
\end{table}

Our ${\cal A}_I$ measurements in the low $s$ region ($0.10<s<8.12$~\gevcccc) yield
\begin{eqnarray}
{\cal A}^{\rm low}_I(B\to K\ellell) & =  -0.58_{-0.37}^{+0.29}\pm0.02 & [2.1\sigma], \nonumber\\
{\cal A}^{\rm low}_I(B\to \Kstar\ellell) & =  -0.25_{-0.17}^{+0.20}\pm0.03 & [1.2\sigma],\nonumber
\end{eqnarray}
where the first uncertainty is statistical and the second is systematic. The
${\cal A}_I$ significances shown in the square brackets include all systematic uncertainties. We estimate the significance
by refitting the data with ${\cal A}_I$ fixed to zero and compute the change
in log likelihood $\sqrt{2 \Delta \ln{\cal L}}$ between the nominal fit and the null hypothesis fit.

\section{Conclusion}
\label{sec:conclusion}

In summary, we have measured total and partial branching fractions, direct $\CP$ asymmetries,
lepton-flavor ratios, and isospin asymmetries in the rare decays $\kll$ using 471 million $B \bar B$ pairs. These results
provide an update to our previous measurements on branching fractions and rate asymmetries excluding the $s < 0.1$~\gevcccc region~\cite{babarrun5}.
The total branching fractions,
${\cal B} (\modekavgll) =  (4.7\pm 0.6\pm 0.2) \times 10^{-7}$ and
${\cal B} (\modekstll)  =  (10.2_{-1.3}^{+1.4}\pm 0.5) \times 10^{-7}$, are measured with precisions of $13\%$ and $14\%$, respectively. The partial branching fractions as a function of $s$ agree well with the SM prediction. 
For $0.10<s<8.12$~\gevcccc, our partial branching fraction results
also allow comparisons with SCET based predictions. \CP asymmetries for both \modekavgll and \modekstll are consistent with zero and the lepton-flavor ratios are consistent with one, both as expected in the SM. The isospin asymmetries at low $s$ values are negative. For $0.10<s<8.12$~\gevcccc we measure
${\cal A}_I (B \ra K \ell^+ \ell^-)  = -0.58^{+0.29}_{-0.37}\pm 0.02$ and ${\cal A}_I (B \ra K^* \ell^+ \ell^-)  = -0.25^{+0.20}_{-0.17}\pm 0.03$. The isospin asymmetries are all consistent with the SM predictions. All results are in good agreement with those of the Belle, CDF, and LHCb experiments.

\section{ACKNOWLEDGMENTS}
\label{sec:acknowledgments}
We are grateful for the 
extraordinary contributions of our \pep2\ colleagues in
achieving the excellent luminosity and machine conditions
that have made this work possible.
The success of this project also relies critically on the 
expertise and dedication of the computing organizations that 
support \babar.
The collaborating institutions wish to thank 
SLAC for its support and the kind hospitality extended to them. 
This work is supported by the
US Department of Energy
and National Science Foundation, the
Natural Sciences and Engineering Research Council (Canada),
the Commissariat \`a l'Energie Atomique and
Institut National de Physique Nucl\'eaire et de Physique des Particules
(France), the
Bundesministerium f\"ur Bildung und Forschung and
Deutsche Forschungsgemeinschaft
(Germany), the
Istituto Nazionale di Fisica Nucleare (Italy),
the Foundation for Fundamental Research on Matter (The Netherlands),
the Research Council of Norway, the
Ministry of Education and Science of the Russian Federation, 
Ministerio de Ciencia e Innovaci\'on (Spain), and the
Science and Technology Facilities Council (United Kingdom).
Individuals have received support from 
the Marie-Curie IEF program (European Union) and the A. P. Sloan Foundation (USA).




\begin{thebibliography}{99}
\bibitem{chargeconj}
  Charge conjugation is implied throughout except as explicitly
      noted.
\end{thebibliography}
\end{document}